%% file: Principal.tex
\documentclass[11pt,a4paper,oneside]{book}
\usepackage[portuges,brazilian]{babel}
\usepackage[pdftex,dvips]{graphicx}

\newcommand{\folhaderosto}[4]{%
  \thispagestyle{empty}
  \begin{singlespace}
    \begin{center}
      \Large\scshape{#1}\\
      \vspace{\stretch{3}}
      \textbf{#2}\\
      \vspace{\stretch{3}}
    \end{center}
    \normalfont\normalsize
    \begin{flushright}
      \parbox[b]{10cm}{#3}
    \end{flushright}
    \begin{center}
      \vspace{\stretch{4}}
      #4
    \end{center}
  \end{singlespace}
  \newpage
}

\newcommand{\dedicatoria}[1]{%
  \thispagestyle{empty}
  \begin{singlespace}
    \hspace{2cm}
    \vspace{\stretch{1}}
    \begin{flushright}
      \parbox[b]{10.5cm}{#1}
    \end{flushright}
    \hspace{2cm}
  \end{singlespace}
  \newpage
}
\usepackage[T1]{fontenc}
\usepackage{setspace}
\usepackage{graphicx}
\usepackage{amsmath}
\usepackage{amsfonts}
\usepackage{amssymb}
\usepackage{amsbsy}
\usepackage{epsfig}
\usepackage{bookman}
\usepackage{fancyhdr}			
\usepackage{latexsym}
\usepackage{cite}
\usepackage{indentfirst}
\usepackage{theorem}
\theoremstyle{plain}

\begin{document}
\input apresentacao
\begin{sloppypar} 
\input Fundamentos1
\input Conclusoes
\end{sloppypar}
\input{bibliografia}

\end{document}

%% file: apresentacao.tex
\pagestyle{myheadings} \thispagestyle{empty} \pagenumbering{roman}

\begin{center}
\vspace{0.2cm} {\Large{\bf Universidade Federal do Espírito Santo\\}}

 \vspace{4cm} {\huge
{\bf  Modelo Acústico Análogo ao Buraco Negro de Schwarzschild\\}}



\vspace{4cm}

{\Large{\bf Júnior Diniz Toniato\\}}

\vspace{4cm}{\large{\bf Orientador: Dr. Sérgio Vitorino de Borba Gonçalves}}
\\
\vspace{2cm} {\bf{Vitória - Espírito Santo\\}} {\bf{2010}}

\end{center}
\newpage


\folhaderosto
{JÚNIOR DINIZ TONIATO}%
{Modelo Acústico Análogo ao Buraco Negro de Schwarzschild}
{Dissertação apresentada ao Programa de Pós-Graduação em Física do Centro de Ciências Exatas da Universidade Federal do Espírito Santo, como requisito parcial para obtenção do Grau de Mestre em  Ciências Físicas.\\
Orientador: Prof. Dr. Sérgio V. B. Gonçalves}
{VITÓRIA\\2010}%


\newpage

\dedicatoria {
\begin{flushright}
  \large{\textit{A todos os meus familiares e amigos que, sem dúvida alguma, foram indispensáveis na minha formação como pessoa.}}
\end{flushright}
}%
\pagebreak

\begin{center}
{\huge Agradecimentos}
\end{center}

\vspace*{1cm}
\begin{sloppypar}

Agradeço ao meu pai, Hélio, que me ensinou a ter caráter em todos os momentos da vida. À minha mãe, Itala, que sempre me educou a ver que tudo, sem qualquer exceção, tem um lado bom e é nele que a gente deve se prender. À minha irmã Kiara, que me provou que irmãos e irmãs podem ser melhores amigos.

Aos meus avôs, tios e primos, que tornam minha família um verdadeiro lar e não somente um amontoado de parentes. Eu sempre serei um reflexo da participação de cada um de vocês na minha vida.

Aos meus amigos agriculanos, Marlos e Verruga (Vinícius), por todas as nossas conversas sobre a questão da vida, o universo e tudo mais. Isso me fez refletir e hoje crer que a Amizade é a única religião e o diálogo é a sua bíblia.

Se for para falar de amizade, agradeço ao Bibício (Fabrício), Gê (Georje), Coelho (Bruno), Gláucio, Crispinga (Tiago), Boi (Marcos) e aos demais amigos que carinhosamente chamo de \textit{Turma do Baco}, pois sempre foram minha segunda família e sempre estiveram do meu lado. Sem falar dos inesquecíveis carnavais que me proporcionaram.

Aos meus tios Rita e Renilton que acreditaram em mim quando eu mais precisava e me acolheram, permitindo assim que eu começasse essa jornada que me trouxe até aqui.

Agradeço aos professores e organizadores do Pré-Vestibular Dandara que fazem um belíssimo trabalho ajudando o próximo sem exigir algo em troca. Em especial aos professores Ramon, Édson e Carlinhos que são os responsáveis por eu ter cursado Física.

Agradeço ao professor Ricardo Berrêdo pelas ótimas aulas, amizade e excelentes discussões, tanto dentro quanto fora das salas de aulas. E por me ensinar que ser um bom físico não é saber falar uma série de palavras complicadas, mas que isso é extremamente divertido!

Aos amigos Dainer, Glauber, Hugo e Rodolfo, pelas conversas, piadas e companheirismo nos quatro anos da graduação. A vocês eu deixo um saudoso Pííí!

Agradeço ao professor Flávio Alvarenga pela orientação desde a Iniciação Científica até o Mestrado. Por ter confiado na minha capacidade, o que gerou ótimos anos de trabalho, aprendizagem e amizade.

Ao professor Sérgio Gonçalves, que sempre foi muito atencioso e prestativo comigo. E também acreditou no meu trabalho sendo importantíssimo para a conclusão deste mestrado.

À CNPq, pelo auxílio financeiro.

Aos companheiros de pós-graduação Raphael e Adriano, que nunca admitiram mas sempre acreditaram que com uma pia e um pouco de água se faz um buraco negro acústico.

À galera do CAFIS e CAMAT por me mostrarem que o IC-1 também tem uma área de lazer, em especial os seus respectivos sofás que permitiam que eu colocasse meu sono em dia. Aos amigos da república Toca do Urso, que estenderam o melhor da UFES para além de seus portões. E aos companheiros do time da Física, foi difícil, mas trouxemos uma medalha pra casa.

\end{sloppypar}

\pagebreak

\vspace*{20cm}
\begin{flushright}
  ``\,Em terra de sací qualquer chutinho é voadora.''
\end{flushright}

\pagebreak

%% file: Fundamentos1.tex
\begin{center}
{\bf Resumo}
\end{center}\vspace{1mm}

Ainda não há uma comprovação experimental que possa validar os resultados obtidos da termodinâmica de buracos negros. Isso porque a radiação emitida pelo buraco negro, prevista pela teoria, é praticamente impossível de ser detectada devido ao baixo valor de sua ordem de grandeza. Na busca por indícios que possam validar a existência dessa radiação, o estudo de modelos análogos a esses objetos tem crescido consideravelmente nas últimas décadas. Eles permitem a idealização de experimentos em laboratórios que seriam impossíveis de serem realizados diretamente nos sistemas gravitacionais.

Um fluido em movimento pode agir sobre o som da mesma forma que os espaço-tempos curvos podem influenciar na trajetória da luz na relatividade geral. Com isso, pode-se descrever a propagação dessas ondas sonoras através de uma métrica efetiva, sob a qual elas seguirão geodésicas nulas. Esta dissertação faz uma revisão destes estudos concentrando-se em uma analogia acústica para um buraco negro de Schwarzschild, demonstrando suas vantagens e limitações quando aplicada para o estudo da teoria de Hawking.

\newpage
\vspace{5mm}

\begin{center}
{\bf Abstract}
\end{center}\vspace{1mm}

There isn't experimental evidences that can validate the results of black holes termodynamics. This because the radiation emitted by the black hole, predicted by theory, it is almost impossible to be detected due to its low value of magnitude. In the search for clues that could validate the existence of this radiation, the study of analog models to those objetcs has grown considerably in recent decades. They allow the idealization of experiments in labs that would be extremely complicate to be done directly with the gravitational systems.

A fluid in moviment can act on the sound the same way that curved space-time can influence on light trajectory in the general relativity. So, one can describe the propagation these sound waves through an effective metric, under wich they will follow null geodesics. This thesis makes a review of these studies focusing in an analogy to Schwarzschild black hole using an acoustic system, showing its advantages and limitations when applied in the study of Hawking's theory.

\tableofcontents

\setcounter{chapter}{0}

\pagestyle{fancy}                       
\fancyhf{}                              
\renewcommand{\chaptermark}[1]{         
  \markboth{\chaptername\ \thechapter.\ #1}{}} %
\fancyfoot[R]{\footnotesize \thepage}   
\fancyhead[L]{\footnotesize \leftmark}
\renewcommand{\headrulewidth}{0.2pt}    
\addtolength{\headheight}{0.5pt}
\makeatletter 
\def\cleardoublepage{\clearpage\if@twoside \ifodd\c@page\else%
   \hbox{}%
    \thispagestyle{empty}
    \newpage%
    \if@twocolumn\hbox{}\newpage\fi\fi\fi} 
\makeatother

\chapter{Introdução}
\setcounter{page}{1}
\pagenumbering{arabic}

A física de buracos negros é uma das áreas de estudos que coloca em prova a teoria da relatividade geral de Einstein pois abrange sistemas com campos gravitacionais tão intensos que permite o surgimento de fenômenos que não são comuns no nosso dia a dia. A primeira idéia de buraco negro foi apresentada em 1783 pelo astrônomo britânico John Michell \cite{michell}, que se baseou na física Newtoniana afirmando que podiam existir estrelas tão compactas, com grande massa, que a velocidade de escape do seu campo gravitacional seria superior a da luz. Naquela época não se nomeava tais objetos de buracos negros e este conceito nem foi muito aceito quando foi implementada a descrição ondulatória da luz, que entre outras coisas, a tornava imune aos efeitos de qualquer campo gravitacional.

No entanto, com a chegada do século XX, e com ele a teoria da Relatividade Geral de Einstein, a gravidade passa a ser interpretada como uma deformação da geometria do espaço-tempo, associando a ele uma curvatura que influencia na trajetória de qualquer partícula. Apenas alguns meses depois da publicação dos trabalhos de Einstein, o físico alemão Schwarzschild \cite{schw1, schw2} obteve sua primeira solução que assumia um objeto esférico massivo na origem das coordenadas. Ele demonstrou a existência de uma singularidade na sua métrica onde toda a matéria poderia vir a se concentrar em um único ponto criando assim uma região de volume nulo e densidade infinita. Uma outra singularidade também surge em uma determinada região, caracterizada pelo que ficou conhecido com o raio de Schwarzschild, que circunda este ponto de densidade infinita. Porém, nem mesmo Schwarzschild tomou estas situações extremas como sendo fisicamente possíveis de ocorrer no universo.

Este quadro mudou em 1939 com o trabalho do físico americano Oppenheimer \cite{opp1, opp2} que desenvolveu uma possível explicação para a natureza dessas singularidades. Ele utilizou das mesmas razões que Chandrazekhar descreveu o colapso gravitacional de uma anã branca em uma estrela de nêutrons para demonstrar que uma estrela de aproximadamente 3 massas solares poderia colapsar sobre ela mesmo até atingir as singularidades de Schwarzschild. Observou-se também que, para um observador externo à estrela, levaria um tempo infinito para que a superfície se reduzisse até o raio de Schwarzschild e, devido a isso, chamavam este colapso de \textit{estrela congelada}, pois era como se sua superfície parasse no tempo quando atingia o raio de Schwarzschild.

Esta nomenclatura foi alterada quando, em 1958, David Finkelstein \cite{fink} introduziu um novo sistema de coordenadas que remove a singularidade da solução no raio de Schwarzschild e verificando que, do ponto de vista de um observador se movendo juntamente com a superfície da estrela, é possível atingir e ultrapassar o raio de Schwarzschild em um tempo finito. A partir daí foi implementado o conceito de horizonte de eventos para essa superfície que anteriormente pensava-se ser singular. Este horizonte na verdade age como um delimitador unidirecional da estrutura causal do espaço-tempo, ou seja, um observador na região interna do horizonte tem acesso a qualquer evento que ocorra na região exterior mas, para um observador externo ao horizonte, tudo o que ocorre na região interior jamais estará acessível a ele. Ressurgia assim o conceito de buraco negro na física, nome este que só começou a ser utilizado em 1967 pelo físico John Weeler durante uma palestra. Este nome ilustra o fato de que estes objetos altamente densos não seriam visíveis para nós por não conseguirem emitir qualquer sinal luminoso devido a existência do horizonte de eventos.

A partir daí muitas características dos buracos negros começaram a ser apresentadas à comunidade científica, bem como outras soluções das equações de Einstein, como o buraco negro de Kerr com rotação e o buraco negro de Reissner-Nordström com carga elétrica (veja alguns exemplos em \cite{townsend}). Na verdade, todas as suas propriedades podem ser descritas pela sua massa, carga e momento angular, já que todo o tipo de massa ou radiação que possa interagir com ele desaparece no horizonte de eventos se tornando inacessível para nós.

Mas um efeito muito interessante, e intrigante até então, pode ocorrer nas proximidades do horizonte de eventos de um buraco negro fazendo com que ele não seja totalmente 'negro': ele emitiria uma certa radiação conhecida como a radiação Hawking \cite{hawking}. Foi o físico britânico, Stephen Hawking, quem fez esta descoberta no início da década de setenta quando aplicou os conceitos da mecânica quântica na física de buracos negros.

No mundo quântico o vácuo não é totalmente vazio, ele contém pares de partículas virtuais que se manifestam como um fóton e um anti-fóton surgindo e se aniquilando muito rapidamente, respeitando o princípio de incerteza de Heinseberg. Acontece que, se estas partículas surgem muito próximo ao horizonte de eventos de um buraco negro, o campo gravitacional ali pode separá-las com a captura do anti-fóton, emitindo um fóton para o infinito. Dessa forma haveria uma radiação sendo emitida no horizonte de eventos de um buraco negro.

Esse fenômeno implica em um gasto de energia pelo buraco negro para separar as partículas do vácuo, o que acaba resultando em um processo de evaporação, ou seja, na diminuição do seu tamanho com o passar do tempo. Mais ainda, a radiação Hawking tem um espectro tal como o emitido por um corpo negro, permitindo que seja associado ao buraco negro uma temperatura conhecida como temperatura Hawking.

Mesmo com toda essa gama de novas informações que veio com a teoria de Hawking, ela nunca teve uma verificação empírica. Isso porque a radiação prevista por ela é extremamente fraca, sendo praticamente impossível de ser detectada com as técnicas usadas pela comunidade científica atualmente. Isso se torna um problema maior se lembrarmos que ainda não existe uma teoria quântica da gravitação. Ou seja, Hawking utilizou de uma física desconhecida como base de sua teoria visto que a transformação da partícula virtual em real no estado de vácuo ocorreria devido a ação da gravidade em um nível quântico.

Muitos trabalhos foram desenvolvidos nessa área desde então, tanto na unificação da relatividade geral com a mecânica quântica quanto na investigação da radiação Hawking numa tentativa de obter informações acerca de sua validade teórica. Um novo caminho foi descoberto no início da década de oitenta pelo físico canadense William George Unruh \cite{unruh}, quando ele notou que ondas sonoras se propagando em um fluido em movimento se comportavam semelhante à luz sob a influência de um campo gravitacional. Se tivermos então um sistema acústico em que o fluido ultrapasse a velocidade do som, criaremos um horizonte de eventos acústico. Isso permitiu a criação de modelos acústicos análogos aos buracos negros da relatividade geral e a idealização de experimentos para os fenômenos previstos pela teoria de Hawking.

Esta dissertação pretende abordar os conceitos básicos mais relevantes para se compreender como esta analogia é feita na mecânica dos fluidos. Essa revisão teórica é feita concentrando-se em um modelo acústico que reproduza a física dos buracos negros de Schwarzschild. A simplicidade proporcionada por essa escolha nos permite ir mais fundo nas discussões quanto a natureza da radiação Hawking.

No capítulo 2 fazemos um estudo em cima da métrica de Schwarzschild e a física de buracos negros que ela envolve. Nos concentramos em discutir todas suas propriedades, desde a caracterização do sistema de coordenadas até a interpretação de suas singularidades. Com isso, analisamos as órbitas para partículas e raios de luz, e descrevemos os sistemas de coordenadas de Finkelstein e Kruskal até obtermos o conceito de buraco negro atual.

O capítulo 3 tem como objetivo obter os mesmos resultados da teoria de Hawking. É feita uma introdução matemática importante sobre horizontes de Killings que nos permite inserir os conceitos de gravidade superficial e temperatura Hawking até obter a formulação de uma teoria termodinâmica de buracos negros.

O capítulo seguinte trata da demonstração de como, em um modelo acústico, a propagação das ondas sonoras em um fluido em movimento pode se assemelhar à
luz se propagando sob influência de um campo gravitacional. Deduzimos a métrica acústica  analisando como esta analogia pode ser estabelecida e quais são seus limites, finalizando com a obtenção de uma expressão para a gravidade superficial acústica, e consequentemente a temperatura Hawking equivalente para o sistema acústico. Neste quarto capítulo fica claro que nosso modelo só consegue imitar a cinemática da métrica de Schwarzschild, sendo assim um modelo limitado quanto à reprodução de uma termodinâmica análoga àquela existente na física de buracos negros.

No quinto e último capítulo, utilizamos um modelo dilatônico para reduzir o buraco negro de Schwarzschild 4D à um sistema 2D (tempo + espaço) e, a partir desta nova configuração geométrica, obtemos um análogo acústico também bidimensional. Isso muda consideravelmente o quadro comentado no parágrafo anterior possibilitando a inclusão da dinâmica na nossa analogia e a definição de grandezas equivalentes à massa e entropia de um buraco negro para o sistema acústico.

\hspace*{\parindent} \setcounter{chapter}{1}
\chapter{A Física de Buracos Negros}
\hspace*{\parindent}

Após a formulação da Teoria da Relatividade Geral, o físico alemão, Karl Schwarzschild obteve a primeira solução exata da equação de Einstein para a gravitação. Esse trabalho, publicado em 13 de janeiro de 1916 e, embora tenha sido relativamente simples por sugerir uma configuração espacial com simetria esférica no vácuo, pode ser considerado o início de toda física de buracos negros conhecida hoje em dia. Iremos estudar neste capítulo essa solução de Schwarzschild \cite{dinverno, hartle, weinberg}, interpretando e analisando suas principais propriedades.

\section{A solução de Schwarzschild}

Como dito anteriormente, a métrica de Schwarzschild é uma solução das equações de campo de Einstein para o vácuo, ou seja, sem fontes. Portanto, ela descreve o campo gravitacional fora de um corpo massivo esférico e sem rotação. Seu elemento de linha, para um grupo particular de coordenadas, é escrito como \cite{dinverno},
\begin{equation}
ds^2=\left(1-\frac{2GM}{c^2r}\right)(cdt)^2-\left(1-\frac{2GM}{c^2r}\right)^{-1}dr^2-r^2\left(d{\theta}^2+{\sin}^2\theta d{\phi}^2 \right).\label{f1}
\end{equation}
Aqui nós adotamos para a assinatura dos coeficientes da métrica $(+,-, -, -)$, tal que $ds^2>0$ é um intervalo do tipo tempo e $ds^2<0$ é um intervalo do tipo espaço. As coordenadas $x^{\mu}=(ct, r, \theta, \phi)$ são chamadas de coordenadas de Schwarzschild e a métrica correspondente $g_{\mu\nu}(x)$ tem as seguintes propriedades:
\begin{itemize}
	\item \emph{A Solução é Estática} - a métrica é independente do tempo $t$. Há um vetor de Killing \textbf{$\xi$} associado com esta simetria sobre deslocamentos na coordenada $t$ que tem componentes
   \begin{equation}
     \xi^{\mu}=(1, 0, 0, 0).\label{f2}
   \end{equation}
  \item \emph{Esfericamente Simétrica} - como já foi dito antes, a métrica de Schwarzschild tem a simetria de uma esfera, e sua invariância sobre rotações no eixo $z$ é representado pelo vetor de Killing com componentes
   \begin{equation}
	  \eta^{\mu}=(0, 0, 0, 1).\label{f3}
   \end{equation}
  \item \emph{Massa $M$} - se $GM/c^2r$ for muito pequeno, o coeficiente de $dr^2$ no elemento de linha (\ref{f1}) pode ser expandido em primeira ordem, tal que
\begin{equation}
	ds^2=\left(1-\frac{2GM}{c^2r}\right)(cdt)^2-\left(1+\frac{2GM}{c^2r}\right)dr^2-r^2\left(d{\theta}^2+{\sin}^2\theta d{\phi}^2 \right).
\end{equation}
E esta é exatamente a forma da métrica de um campo gravitacional fraco e estático com um potencial gravitacional \cite{hartle}, dado por
\begin{equation}
	\Phi=-\frac{GM}{r}.
\end{equation}
Isto nos leva a identificar a constante $M$ na métrica Schwarzschild como sendo a massa total da fonte de curvatura do espaço-tempo. Portanto, a geometria fora de uma fonte esfericamente simétrica é caracterizada por uma única grandeza (sua massa total) e não como esta massa está radialmente distribuída dentro da fonte. Por conveniência nós definimos uma \textit{massa geométrica} $m$ que tem dimensão de comprimento,
\begin{equation}
	m\equiv \frac{GM}{c^2}. \label{f4}
\end{equation}
\end{itemize}
Levando em consideração os ítens descritos acima, podemos reescrever a equação (\ref{f1}) como
\begin{equation}
	ds^2=\left(1-\frac{2m}{r}\right)(cdt)^2-\left(1-\frac{2m}{r}\right)^{-1}dr^2-r^2(d\theta^2-sen^2\theta d\phi^2). \label{f5}
\end{equation}

\subsection{Caracterização das coordenadas}
Uma métrica pode ser representada por diversos sistemas de coordenadas, porém, se ela possui certas simetrias então existirá coordenadas preferenciais adaptadas conforme estas simetrias.

Observando as componentes diagonais da métrica de Schwarzschild concluímos que, para $r>2m$, $x^0=ct$ é tipo tempo e $x^1=r$, $x^2=\theta$ e $x^3=\phi$ são tipo espaço. Como a métrica é independente de $t$ e não há termos cruzados com $dt$, ela é estática e $t$ é identificado como uma coordenada que representa o tempo. A coordenada $r$ é um parâmetro radial com a propriedade de que a 2-esfera $t=constante$ e $r=constante$ tem o elemento de linha
\begin{equation}
	ds^2=-r^2(d\theta^2 + \sin^2\theta d\phi),
\end{equation}
de onde segue que a área da superfície dessa 2-esfera é $4\pi r^2$. Finalmente, as coordenadas $\theta$ e $\phi$ são os ângulos usuais das coordenadas esféricas, que são invariantemente definidas pela simetria esférica. Portanto, as coordenadas de Schwarzschild são coordenadas canônicas definidas invariantemente pelas simetrias presentes na métrica.

A variação radial infinitesimal da distância própria é obtida da métrica fazendo $dt=d\theta=d\phi=0$. Essa distância própria não é $dr$, mas sim $dr/(1-2m/r)^{1/2}$. A coordenada $r$ portanto não mede distância própria.
De forma similar, a coordenada $t$ também não é o tempo próprio. A variação infinitesimal no tempo próprio é relacionada com a variação infinitesimal da coordenada $t$ tal que $ds=cd\tau$. Fazendo então $dr=d\theta=d\phi=0$ no elemento de linha (\ref{f1}), teremos
\begin{equation}
	 d\tau=\left(1 - \frac{2m}{r}\right)^{1/2}dt.\label{f6}
\end{equation}
Note que $d\tau=dt$  quando $r\rightarrow+\infty$ tal que a coordenada temporal $t$ é interpretada como sendo o tempo próprio quando medida a uma distância infinita da fonte do campo gravitacional.

A consequência desta diferença na medida do tempo localmente e no infinito é que a radiação enviada de uma certa posição de raio $r$ sofre um desvio para o vermelho quando é recebida por um outro observador bem distante. Como o comprimento de onda da radiação é proporcional ao período da vibração, a equação (\ref{f6}) nos diz que
 \begin{equation}
	 \lambda=\left(1-\frac{2m}{r}\right)^{1/2}\lambda_{\infty},
 \end{equation}
para a relação entre o comprimento de onda $\lambda$ da radiação emitida em $r$ e o comprimento de onda $\lambda_{\infty}$ recebido no infinito. Para o desvio para o vermelho $z$, definido por $z=(\lambda_{\infty}-\lambda)/\lambda$, \cite{Raine} temos que
\begin{equation}
	1+z=\left(1 - \frac{2m}{r}\right)^{-1/2}.\label{f7}
\end{equation}

Considere agora um caso onde $r<2m$. Nesta região, as coordenadas $r$ e $t$ invertem seus papéis. Por exemplo, para um corpo em $r<2m$, com $\theta$, $\phi$ e $t$ constantes, temos $d\tau^2=g_{11}dr^2>0$, então $dr$ é um intervalo \textit{tipo tempo} nessa região. Isto significa que o corpo não poderá permanecer em repouso, ele irá cair até atingir a singularidade $r=0$. Mais ainda, como os coeficientes da métrica dependem do tempo ($r$ nesse caso), ela não é mais estática nessa região. No caso da coordenada $t$, ela passa a desempenhar um papel de coordenada espacial.

\section{Singularidades}
De acordo com a métrica (\ref{f1}) vemos que a métrica se torna degenerada em dois pontos específicos: $r=2m$, que torna $g_{11}$ divergente, e $r=0$ onde $g_{00}$ é divergente. O primeiro ponto é conhecido como o \textit{raio de Schwarzschild} ($r_s$) e identifica uma singularidade de coordenadas. Uma singularidade deste tipo pode ser removida com uma transformação de coordenadas adequada, sendo assim considerada como um resultado da matemática utilizada para descrever o espaço-tempo em questão.

Isso pode ser verificado através do invariante escalar do tensor de Riemann \cite{dinverno}, calculado para a métrica de Schwarzschild
 \begin{equation}
    R_{abcd}R^{abcd}=\frac{48m^2}{r^6},
 \end{equation}
que é finito para $r=r_s$. Por ser um escalar, seu valor é o mesmo em qualquer sistema de coordenadas. Porém, esse valor diverge para $r=0$ o que implica que a singularidade neste tempo não é removível como a anterior. Esta então é uma singularidade real e um problema físico proveniente da própria métrica de Schwarzschild. É importante notar também que na hiperfície $r=r_s$ há um desvio para o vermelho infinito, que pode ser visto pela equação (\ref{f7}).

Se considerarmos casos onde a fonte de gravitação tenha um raio $R>r_s$ nós não teríamos problema algum com essa singularidade em $r=2m$. Isso porque a métrica de Schwarzschild só é válida para a região exterior a fonte do campo gravitacional e nesse caso este ponto singular não estaria dentro do domínio da solução. No interior do corpo massivo o tensor momento-energia na equação de Einstein não será nulo, ou seja, uma nova solução deve ser obtida para essa região. No entanto, a principal questão é verificar o que acontece quando a fonte tem um raio menor que o raio de Schwarzschild. Nesse caso a solução continuará sendo válida na região exterior ao corpo massivo, mas passamos a ter um espaço-tempo dividido em duas regiões desconexas, $0<r<r_s$ e $r>r_s$.

\section{Órbitas na métrica de Schwarschild}
Vamos agora estudar as órbitas descritas por partículas de teste na geometria de Schwarzschild. Considerar as simetrias esférica e temporal torna esta análise mais simples. E para simplificar ainda mais faremos $c=1$.

\subsection{Constantes do movimento}
As simetrias são extremamente úteis pois a partir delas podemos obter algumas grandezas que permanecem constantes durante o movimento da partícula. Para o nosso caso particular, como a métrica é independente de $t$ e $\phi$, as quantidades $\boldsymbol{\xi}\cdot \bf{u}$ e $\boldsymbol{\eta} \cdot \bf{u}$ são conservadas, onde $\bf{u}$ é a quadri-velocidade da partícula cuja as componentes são $u^{\mu}=dx^{\mu}/d\tau$, sendo $x^{\mu}(\tau)$ o caminho seguido por essa partícula. Os vetores de Killing $\boldsymbol{\xi}$ e $\boldsymbol{\eta}$ são dados por (\ref{f2}) e (\ref{f3}). Assim, temos que
\begin{align}
	\boldsymbol{\xi}\cdot {\bf u}=g_{\mu\nu}&\xi^{\mu}u^{\nu}=g_{00}u^0=E,\nonumber \\	
   \nonumber \\
   &\left(1-\frac{2m}{r}\right)\frac{dt}{d\tau}=E,\label{f8}
\end{align}
onde $E$ é uma constante e $m$ é dado pela equação (\ref{f4}). Se considerarmos uma partícula de massa unitária, $u^{\mu}$ é o seu quadri-momento. Portanto, a equação (\ref{f8}) representa a conservação da componente temporal do quadri-momento, isto é, a conservação da energia. A quantidade $E$ então, é a energia relativística por unidade de massa da partícula.

Analogamente para a simetria na direção azimutal, teremos
\begin{align}
	 \boldsymbol{\eta}\cdot {\bf u}= g_{\mu\nu}\eta^{\mu}u^{\nu}&=g_{33}u^3=-L,\nonumber \\
	\nonumber \\
	&r^2\sin^2{\theta}\frac{d\phi}{d\tau}=L, \label{f9}
\end{align}
com $L$ sendo uma constante também. De maneira análoga, interpretamos a conservação da componente $\phi$ do quadri-momento como a conservação do momento angular, com $L$ sendo o momento angular por unidade de massa da partícula.

A conservação do momento angular implica que a órbita esta contida em um único plano e para simplificar as contas escolhemos uma órbita no plano equatorial, onde $\theta=\pi/2$, e com isso $d\theta=0$ e $sen~\theta=1$. Dividindo então a equação (\ref{f5}) por $d\tau^2$ teremos
\begin{equation}
	1=\left(1-\frac{2m}{r}\right)\left(\frac{dt}{d\tau}\right)^2 - \left(1-\frac{2m}{r}\right)^{-1}\left(\frac{dr}{d\tau}\right)^2 - r^2\left(\frac{d\phi}{d\tau}\right)^2. \label{f10}
\end{equation}
Utilizando as equações (\ref{f8}) e (\ref{f9}) escrevemos
\begin{equation}
	\left(\frac{dr}{d\tau}\right)^2=E^2 - \left(1+\frac{L^2}{r^2}\right)\left(1-\frac{2m}{r}\right)= E^2-V^{2}_{ef}(r). \label{f11}
\end{equation}

\begin{figure}[htpb]
    \centering
		\includegraphics[scale=0.4]{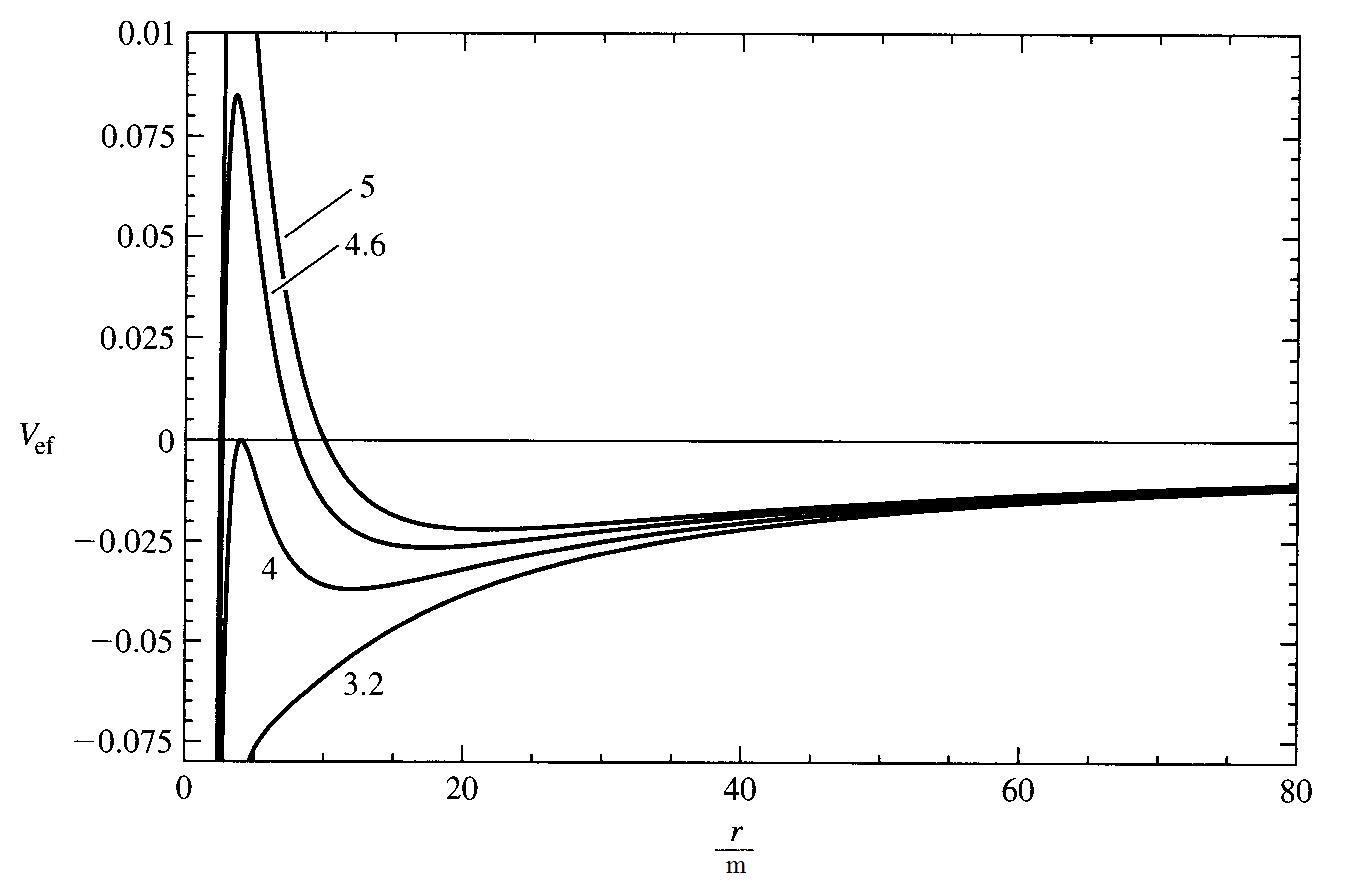}
	\label{fig1}
	\caption{Potencial efetivo para diferente valores de $L$, plotado como $1/2[V^2_{ef}-1]$ em função de $r/m$. Os números indicados em cada curva são os valores de $L/m$. }
\end{figure}

A quantidade $V_{ef}(r)$ é conhecido como o potencial efetivo e pode fornecer informações interessantes sobre órbitas na métrica de Schwarzschild. A figura~\ref{fig1} mostra o potencial efetivo em função de $r/m$ e nota-se que há órbitas fechadas e abertas, mas nada pode ser afirmado se elas são elípticas ou hiperbólicas. A característica mais intrigante é que nenhum valor do momento angular pode manter uma órbita com energia suficiente fora da região $r<2m$. Pior do que isso, para partículas com pouco momento angular em relação a massa central, na verdade $L<2\sqrt{3}m$, todas as órbitas acabarão dentro da região $r<2m$. Isso é bem diferente do que é obtido para a mecânica Newtoniana, onde qualquer momento angular previne uma partícula de atingir o ponto $r=0$. Esta é a primeira evidência que vemos de que para um corpo esférico com raio menor que o raio de Schwarzschild coisas peculiares podem acontecer para pequenos valores de $r$. Na verdade, isso mostra que, nenhum objeto massivo é capaz de sair da região $r<2m$, uma vez que tenha entrado nela.

\subsection{Queda livre radial}
Considere uma partícula de massa de repouso $m_0$ em queda livre radial, seu quadri-momento é $(p^{\alpha})=(m_0u^0, m_0u^1, 0, 0)$ para um movimento puramente radial. De (\ref{f8}) temos que
\begin{equation}
	 p^0=m_0\frac{dt}{d\tau}=\frac{m_0E}{\left(1-\frac{2m}{r}\right)},\label{f12}
\end{equation}
e, usando $m_0^2=g_{\mu\nu}p^{\mu}p^{\nu}$, obtemos
\begin{equation}
	 p^1=\pm m_0\left(E^2-1+\frac{2m}{r}\right)^{1/2}.\label{f13}
\end{equation}

Já vimos anteriormente que, para um observador situado bem distante da fonte gravitacional, a energia total por unidade de massa de uma partícula em queda livre radial tem um valor constante $E$. Por comparação, a energia $\epsilon$ medida por um observador local, mantido em uma distância $r$ fixa, com quadri-velocidade $(u^{\mu})=(dt/d\tau, 0, 0, 0)$, é dada por $u^{\mu}p_{\mu}$, a projeção do quadri-momento da partícula ao longo de $u^{\mu}$. Para um observador em repouso no espaço-tempo de Schwarzschild, da métrica temos que
\begin{equation}
	d\tau^2=(1-2m/r)dt^2,
\end{equation}
então
\begin{equation}
	(u^{\mu})=([1-2m/r]^{-1/2}, 0, 0, 0),
\end{equation}
e assim
\begin{equation}
	 \epsilon=\frac{m_0E}{(1-\frac{2m}{r})^{1/2}}.\label{f14}
\end{equation}
Segue que a energia medida por um observador local estacionário aumenta com a redução do valor de $r$. A equação (\ref{f14}) mostra que as duas energias são relacionadas por um fator de desvio para o vermelho, o que é justamente o requerido pela conservação da energia.

\subsection{Órbitas circulares}
Uma partícula em órbita circular tem sua coordenada $r$ constante ao longo do tempo, ou seja, não há velocidade nem aceleração radial. Portanto, as relações $dr/d\tau=0$ e $d^2r/d\tau^2=0$ devem ser satisfeitas. A primeira condição, juntamente com a equação (\ref{f11}), informa que $E=V_{ef}(r)$. Da equação radial (\ref{f11}) também obtemos que
\begin{equation}
 \frac{d^2r}{d\tau^2}=-\frac{1}{2}\frac{d}{dr}V^2_{ef}(r).
\end{equation}
Então, utilizando a condição de que o movimento da partícula não é acelerado na coordenada radial, temos
\begin{equation}
 \frac{dV^2_{ef}}{dr}=2V_{ef}\frac{dV_{ef}}{dr}=0,
\end{equation}
logo, $dV_{ef}/dr=0$. Portanto, órbitas circulares somente são possíveis em pontos extremos do potencial efetivo.

Usando a equação (\ref{f11}) novamente, podemos obter estes pontos extremos do potencial efetivo,
\begin{equation}
	 r=\frac{L^2}{2m}\pm\frac{1}{2}\sqrt{\frac{L^4}{m^2}-12L^2}.\label{f15}
\end{equation}
Desta equação vemos que há duas soluções para $L^2>12m^2$, o sinal negativo corresponderá a um máximo desse potencial, o que é um ponto instável, e o sinal positivo indicará um mínimo, que será estável. Quando $L^2=12m^2$ há somente uma órbita circular e esta será ligeiramente estável, pois $d^2V_{ef}/dr^2=0$, e é a órbita 'estável' existente de menor raio, $r=6m$. Como esta órbita é somente ligeiramente estável qualquer perturbação, por menor que seja, levará a partícula até $r=2m$.

\section{Órbitas de raios de luz}
O cálculo de órbitas de raios de luz na geometria de Schwarzschild é semelhante ao cálculo de órbitas de partículas. As quantidades $\boldsymbol{\xi}\cdot\bf{u}$ e $\boldsymbol{\eta}\cdot\bf{u}$ são conservadas, mas estamos tratando agora de fótons, partículas que têm massa de repouso nula. Isso vai influenciar diretamente na definição da quadri-velocidade.

Os fótons seguem geodésicas nulas por isso a curva $x^{\alpha}$ não pode ser parametrizada em função do tempo próprio $\tau$. Na ausência de qualquer campo gravitacional os raios de luz seguem trajetórias tal que $x=t$ e esta curva pode ser parametrizada como
\begin{equation}
x^{\alpha}=u^{\alpha}\lambda,
\end{equation}
onde $\lambda$ é o parâmetro da curva e $u^{\alpha}=(1, 1, 0, 0)$. Com essa parametrização temos que $u^{\alpha}=dx^{\alpha}/d\lambda$, ou seja, $\bf{u}$ é o quadri-vetor tangente à curva e pode ser usado exatamente como no caso das partículas para definir as órbitas de raios de luz. Assim, teremos que, tal como foi feito em (\ref{f8}) e (\ref{f9}),
\begin{equation}
	\left(1-\frac{2m}{r}\right)\frac{dt}{d\lambda}=E_f, \label{f16}
	\end{equation}
e
\begin{equation}
	r^2\frac{d\phi}{d\lambda}=L_f, \label{f17}
\end{equation}
onde $E_f$ e $L_f$ são constantes. Pode-se também normalizar $\lambda$ tal que $\bf{u}$ coincida com o quadri-momento do fóton movendo-se ao longo da geodésica nula. Assim, as equações acima representam a conservação da energia e do momento angular para um fóton. Note que, pela própria definição, $\bf{u}$ é um vetor nulo. Dessa forma,
\begin{align}
  \bf{u}\cdot\bf{u} &= 0, \nonumber \\
	\nonumber \\
	\left(\frac{dr}{d\lambda}\right)^2=E_f^2 - \frac{L_f^2}{r^2}&\left(1-\frac{2m}{r}\right)=E_f^2 - V_f^2. \label{f18}
\end{align}
Analogamente ao caso com partículas massivas obtemos um potencial efetivo, $V_f(r)$, que nos ajudará a classificar melhor as órbitas de raios de luz. Na verdade, a única diferença entre os dois casos é o valor do produto escalar do vetor tangente ($0$ para luz e $1$ para partículas). Este nível de semelhança ocorre graças à parametrização que fizemos. Para diferentes parâmetros a definição de $\bf{u}$ pode mudar, mas ele continua sendo nulo.

O máximo do potencial efetivo ocorre quando $r=3m$ independentemente do valor de $L_f$. Portanto esta órbita é circular mas, devido ao fato de ser um máximo do potencial, ela é instável. Não há um mínimo para $V_f$, logo não há órbitas circulares estáveis para fótons. Isso nos diz que, na teoria um corpo esférico pode prender a luz em uma órbita circular, mas na prática isso nunca acontece.

Devido à liberdade na normalização do parâmetro afim, $\lambda$, se o multiplicarmos por uma constante qualquer, a norma de $\bf{u}$ e a equação (\ref{f18}) não mudarão. Mas os valores de $E_f$ e $L_f$ mudarão. Portanto, diferente do caso de partículas, as propriedades físicas das órbitas de raios de luz podem depender somente da razão entre $E_f$ e $L_f$. Definimos então o escalar $b\equiv{L_f/E_f}$ que contém as informações acerca do comportamento das órbitas. Para entender melhor o papel de $b$ considere uma órbita de luz no infinito, onde o espaço-tempo é plano. Podemos introduzir as coordenadas cartesianas e tomar este raio de luz em questão como se propagando paralelamente ao eixo $x$, a uma distância $d$ dele. Para $r>>2m$, a quantidade $b$ é
\begin{equation}
	b\equiv \left|\frac{L_f}{E_f}\right| \approx \frac{r^2d\phi/d\lambda}{dt/d\lambda}=r^2\frac{d\phi}{dt}.
\end{equation}
Para $r$ muito grande nós temos $\phi \approx d/r$ e $dr/dt \approx -1$, resultando em
\begin{equation}
	\frac{d\phi}{dt}=\frac{d\phi}{dr}\frac{dr}{dt}=\frac{d}{r^2}.
\end{equation}
Portanto
\begin{equation}
	b=d,
\end{equation}
e a constante $b$ representa então o \textit{parâmetro de impacto} de um raio de luz.

No ponto de máximo do potencial efetivo temos $V_f^2(3m)=L_f^2/27m^2$, de onde podemos escrever que
\begin{equation}
	V_{f~max}^2=\frac{b^2}{27m^2}E_f^2. \label{f19}
\end{equation}
Assim, a característica da órbita dos raios de luz depende de $b^2$ ser maior ou menor que $27m^2$. Considerando órbitas vindas do infinito, se $b^2<27m^2$, a luz simplesmente será desviada e seguirá novamente para o infinito. Se $b^2>27m^2$ a luz circulará até ser capturada.

Se considerarmos órbitas que se iniciam em pequenos raios, algo entre $r=2m$ e $r=3m$, no caso onde $b^2>27m^2$ a luz consegue escapar. Quando $b^2<27m^2$ há um ponto onde ela muda de sentido e é capturada pela fonte gravitacional. Isso indica que, se o feixe de luz começa sua órbita com momento angular pequeno, isto é, suficientemente próximo da direção radial, então ela escapará do campo gravitacional.

\subsection{Propagação radial da luz}
O quadri-momento de um raio de luz se propagando radialmente será 
\begin{equation}
	(p^{\alpha})=(dt/d\lambda, \pm~dr/d\lambda, 0, 0).
\end{equation}
Portanto, de (\ref{f16}),
\begin{equation}
	 p^0=\frac{E_{\infty}}{(1-2m/r)},
\end{equation}
onde $E_{\infty}$ é a energia do fóton no infinito. Como $L_f$ é nulo, $p^1=E_{\infty}$, de modo que
\begin{equation}
	 (p^{\alpha})=\left(\frac{E_{\infty}}{(1-2m/r}, \pm E_{\infty}, 0, 0\right). \label{f20}
\end{equation}

\section{Colapso Gravitacional}
A partir do momento em que uma estrela se forma ela atinge um estado estático onde a energia perdida na radiação é balanceada com aquela produzida pelas reações nucleares em seu interior, em especial a queima de hidrogênio. Esse é o estado em que se encontra hoje o nosso Sol, por exemplo. Porém, isso não dura para sempre, a evolução atingirá um estado onde não haverá mais reações termonucleares ocorrendo, chegando a estrela a um estado final de equilíbrio como as anãs brancas ou estrelas de neutrons. No entanto, para casos onde a massa é muito grande (da ordem de 3 ou mais massas solares), a relatividade geral prevê que a pressão interna pode não suportar a atração gravitacional e a superfície dessa estrela pode colapsar.

Vamos assumir que a pressão de suporte foi removida completamente e que a superfície está em queda livre. Especificamente, estamos interessados em saber quanto tempo leva, do ponto de vista de um observador situado a uma distância $r>>2m$ fixa, para a superfície atingir o horizonte de eventos em $r=2m$.

De acordo com a equação (\ref{f13}) temos que
\begin{equation}
	 \left(\frac{dr}{d\tau}\right)^2=E^2-1+\frac{2m}{r}.
\end{equation}
Supondo o início do movimento da partícula $m_0$ em repouso no infinito, então $E^2=1$ e
\begin{equation}
	\left(\frac{dr}{d\tau}\right)^2=\frac{2m}{r}. \label{f21}
\end{equation}
Entretanto, como estamos observando a uma distância muito grande, devemos usar como \textit{nosso} tempo próprio a coordenada $t$. Então, como
\begin{equation}
	 \frac{dt}{d\tau}=\frac{1}{1-2m/r},
\end{equation}
teremos
\begin{equation}
	 \frac{dr}{dt}=\pm \sqrt{\frac{2m}{r}}\left(1-\frac{2m}{r}\right), \label{f22}
\end{equation}
descrevendo a queda da superfície vista de um observador distante. Integramos a equação acima de $r'=r$ até $r'=R$, onde $R$ é muito próximo do horizonte de eventos (usando $r'$ como variável de integração).
\begin{equation}
	 t=\int_r^R{\left(\frac{r'}{2m}\right)^{1/2}\frac{r'dr'}{r'-2m}}. \label{f23}
\end{equation}
Mas a maior contribuição ocorre próximo do horizonte, então fazendo $r'=2m+\epsilon$ e expandindo em potências de $\epsilon$,
\begin{equation}
	 \left(\frac{r'}{2m}\right)^{1/2}\frac{r'}{r'-2m}\approx \frac{2m}{\epsilon},
\end{equation}
e assim
\begin{equation}
	 t\approx -2m \ln{\left(\frac{r}{2m}-1\right)}. \label{f24}
\end{equation}
Este é o tempo gasto para que a superfície sofra um colapso até as proximidades de $r=2m$. Note que este tempo tende ao infinito quando $r\rightarrow{2m}$.

\subsection{Tempo de viagem da luz}
Mas o observador só consegue enxergar este evento quando os fótons provenientes da superfície da estrela chegarem até ele. Por isso devemos adicionar este tempo de viagem da luz a equação (\ref{f24}).

Para a propagação radial da luz, $d\tau^2=0$, e com $d\theta=0$ e $d\phi=0$ obtém-se
\begin{equation}
	\frac{dt}{dr}=\frac{1}{1-2m/r}. \label{f24a}
\end{equation}
Integrando (\ref{f24a}),
\begin{equation}
	t'=\int_r^R{\frac{r'dr'}{r'-2m}}\approx -2m \ln\left(\frac{r}{2m}-1\right),
\end{equation}
onde usou-se outra vez a contribuição maior nas proximidades do horizonte. O observador vê a superfície atingindo o horizonte no tempo $T=t+t'$, ou seja,
\begin{equation}
	T=-4m\ln\left(\frac{r}{2m}-1\right). \label{f25}
\end{equation}
Isto continua divergindo quando $r=2m$. Portanto, para o observador distante da estrela, a queda nunca termina. Ele nunca verá a superfície (ou um objeto qualquer) atingir o horizonte de eventos.

\subsection{Observador em queda junto com a superfície}
Da equação (\ref{f21}), em termos do tempo próprio de um observador em queda livre, temos
\begin{equation}
	 \tau=-\int_R^{2m}{\left(\frac{r}{2m}\right)^{1/2}dr}=\frac{4}{3}m\left[\left(\frac{R}{2m}\right)^{3/2}-1\right] \label{f26}
\end{equation}
representando o tempo de queda de $r=R$ até $r=2m$. Portanto, o observador atinge o horizonte de eventos e o ultrapassa em um tempo finito. Isso deixa claro que o sistemas de coordenadas de Schwarzschild não é apropriado para descrever todo os espaço-tempo ao redor de uma massa esférica. Devemos utilizar um novo sistemas de coordenadas onde a singularidade em $r=2m$ seja devidamente removida.

\section{Diferentes sistemas de coordenadas}

Vimos anteriormente que há um grande problema na métrica de Schwarzschild em relação ao ponto $r=2m$. Não tem como conectarmos as duas regiões que esse sistema de coordenadas descreve, pois a própria métrica diverge nesse ponto. Para isso então, utilizamos alguns sistemas de coordenadas que possibilitam descrever de forma contínua todo o espaço-tempo exterior ao corpo massivo.

\subsection{Coordenadas de Eddington-Finkelstein}
Iniciamos procurando por uma equação nas coordenadas de Schwarzschild que descreva raios de luz. Isto pode ser obtido integrando $d\tau^2=0$ para uma propagação radial. De (\ref{f5}), temos que
\begin{equation}
	dt^2=\frac{dr^2}{(1-2m/r)^2}\equiv dr^2_*, \label{f27}
\end{equation}
onde temos definido uma nova coordenada radial $r_*$ de tal forma que os raios de luz são descritos por $d(t\pm r_*)=0$ ou $t \pm r_*=cte$. Integrando esta equação obtemos
\begin{equation}
	r_*=r+2m~log\left|\frac{r-2m}{2m}\right|. \label{f28}
\end{equation}
Usaremos $v=t+r_*$ como a nova coordenada temporal, substituindo $t$. Com isso, reescrevemos a métrica de Schwarzschild e obtemos
\begin{equation}
	 d\tau^2=\left(1-\frac{2m}{r}\right)dv^2-2dvdr-r^2d\tilde{\omega}^2. \label{f29}
\end{equation}
Com essa nova métrica, podemos obter algumas conclusões acerca da região interior ao horizonte de eventos, pois agora ela não é mais singular em $r=2m$. Para analisar essas propriedades reescrevemos a equação (\ref{f29}) como
\begin{equation}
	 2drdv=-\left[d\tau^2 + \left(\frac{2m}{r}-1\right)dv^2 +r^2d\tilde{\omega}^2\right]. \label{f30}
\end{equation}
Para um deslocamento tipo tempo ou nulo, $d\tau^2\geq0$, e se $r<2m$ o lado direito da equação acima será negativo. Lembrando que, feixes de luz se propagando radialmente, são descritos por $dv=d(t+r_*)=0$, e a linha do tempo de qualquer objeto material deve estar contida no interior do cone de luz, o deslocamento $dv$ será sempre $dv\geq0$. Assim, para fazer o lado esquerdo da equação (\ref{f30}) ser negativo temos que ter $dr<0$. Isto implica que a singularidade $r=0$ é um futuro inevitável para qualquer deslocamento na região $r<2m$.

O horizonte de eventos separa todos os eventos em seu interior do mundo exterior a ele. Um explorador que supostamente atravessasse este horizonte nunca conseguiria voltar de sua viagem e nem sequer poderia enviar qualquer informação sobre suas experiências para a região exterior. E muito pior do que isso, ele seria incapaz de evitar sua própria destruição na singularidade em $r=0$. Portanto, começa a ficar clara e mais explícita a noção de buraco negro: corpo esférico massivo com raio menor que o raio de Schwarzschild. O horizonte de eventos age então como o delimitador da região de onde nem a luz consegue escapar.

As coordenadas de Eddington-Finkelstein que usamos foram baseadas em raios nulos que entram na região $r=2m$ no futuro (para $v=cte$, $t$ aumenta à medida que $r_*$ diminui). Se considerarmos $u=t-r_*$ como coordenada temporal, descrevemos raios se afastando do horizonte ($t$ aumenta junto com $r_*$, mantendo $u$ constante), que na métrica de Schwarzschild retornará
\begin{equation}
	d\tau^2=\left(1-\frac{2m}{r}\right)du^2 +2drdu - r^2d\tilde{\omega}^2.
\end{equation}
Isto descreve um espaço-tempo que é o inverso de um buraco negro, onde todo futuro de deslocamentos nulos ou tipo tempo emergem de $r<2m$ para o espaço ao redor. Isto é chamado de buraco branco. O importante é notar que, tanto buracos negros quanto buracos brancos são soluções matemáticas das equações de vácuo de Einstein, mas isto não significa que ambos sejam fisicamente possíveis de existir.

\subsection{Coordenadas de Kruskal}

As coordenadas de Eddington-Finkelstein descrevem um espaço-tempo contendo um buraco negro ou um buraco branco, mas nunca as duas simultaneamente. Seria interessante poder olhar para essas duas regiões em uma mesma métrica. Isto é feito introduzindo as coordenadas de Kruskal, que também nos fornece informações adicionais para entender melhor a física existente no horizonte de eventos.

Retornando a métrica de Schwarzschild, fazendo uso da equação (\ref{f27}), obtemos
\begin{equation}
	d\tau^2=\left(1-\frac{2m}{r}\right)(dt^2-dr_*^2)-r^2 d \tilde{\omega},
\end{equation}
e utilizando as duas coordenadas nulas $u=t-r_*$ e $v=t+r_*$, temos
\begin{equation}
	d\tau^2=\left(1-\frac{2m}{r}\right)dudv-r^2d\tilde{\omega}.
\end{equation}
Agora definimos
\begin{eqnarray}
	U&=&-e^{\frac{-u}{4m}},\label{f32}\\
	V&=&e^{\frac{v}{4m}},\nonumber
\end{eqnarray}
o que nos leva à forma
\begin{equation}
	d\tau^2=\frac{32m^2}{r}e^{-r/2m}dUdV-r^2d\tilde{\omega}, \label{f33}
\end{equation}
onde consideramos $r$ como sendo uma função de $U$ e $V$, tal que
\begin{equation}
	VU=-\left(\frac{r}{2m}-1\right)e^{r/2m}, \label{f34}
\end{equation}
obtido com uso da equação (\ref{f28}). A equação (\ref{f33}) é a métrica de Schwarzschild nas coordenadas de Kruskal $(U, V, \theta, \phi)$. Inicialmente a métrica é definida para $U<0$ e $V>0$, mas através de um extensão analítica, podemos estender para $U>0$ e $V<0$. Note que $r=2m$ corresponde a $UV=0$, isto é, $U=0$ ou $V=0$.

Com isso, é conveniente traçar linhas de $U$ e $V$ constantes para obter o diagrama de espaço-tempo de Kruskal. Há quatro regiões distintas neste diagrama, cada uma definida de acordo com os sinais de $U$ e $V$.

\begin{figure}[htpb]
    \centering
		\includegraphics[scale=0.8]{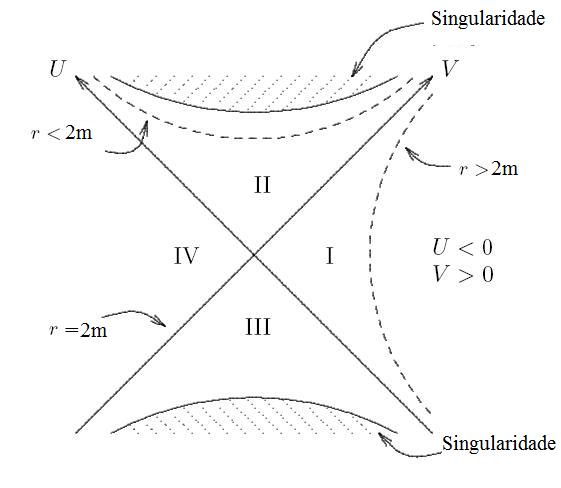}
	\label{fig2}
	\caption{Diagrama de Kruskal.}
\end{figure}

As regiões I e II correspondem ao espaço-tempo descrito pelas coordenadas de Eddington-Finkelstein para buracos negros. As regiões I e III são referentes ao espaço-tempo de um buraco branco de Eddigton-Finkesltein. Mas agora notamos que surge uma nova parte no espaço-tempo que só conseguíamos ver com as coordenadas de Kruskal, a região IV. As equações matemáticas então mostram a existência de um novo universo com a região IV, mas isto não quer dizer que ele realmente exista fisicamente. Ele é geometricamente idêntico ao espaço assintoticamente plano de Schwarzschild, a região I.


\chapter{Termodinâmica de um Buraco Negro}
Na década de setenta, o físico britânico Stephen Hawking publicou seu trabalho sobre buracos negros que se comportam como corpos negros absorvendo e emitindo radiação \cite{hawking}. Esta radiação emitida tem um espectro de potência de Planck que nos permite associar ao buraco negro uma certa temperatura.

Basicamente o que Hawking fez foi considerar os efeitos quânticos de criação de partículas nas proximidades do horizonte de eventos. No vácuo, as flutuações quânticas se manifestam com o surgimento de um par de fóton e anti-fóton que se aniquilam em um intervalo de tempo que não viole o princípio da incerteza de Heisenberg \cite{landau2}. O que ocorre é que, se estes pares surgem muito próximo do horizonte, um dos fótons pode ser capturado pelo buraco negro enquanto o outro seria emitido para o infinito. Após essa descoberta teórica, Hawking avançou ainda mais estabelecendo relações entre as variações de massa e área com a temperatura, tal como é feito na termodinâmica clássica entre energia, temperatura e entropia, originando a chamada termodinâmica de buracos negros. Nas seções seguintes iremos, através de conceitos basicamente matemáticos, reproduzir os resultados de Hawking.

\section{Hipersuperfícies nulas}
Se tomarmos a superfície $S=r-2m$ nas coordenadas de Schwarzschild e calcularmos o seu vetor normal $l^{\mu}$, teremos
\begin{equation}
	l^{\mu}=f(x)g^{\mu\nu}\partial_{\nu}S=f(x)\left(1-\frac{2m}{r}\right),
\end{equation}
onde $f(x)$ é uma função arbitrária. Tomando a norma desse vetor,
\begin{equation}
	l^2=g_{\mu\nu}l^{\mu}l^{\nu}=f^2(x)\left(1-\frac{2m}{r}\right),
\end{equation}
notamos que ela é nula para $r=2m$. Então, o vetor normal ao horizonte de eventos tem norma nula, o que define uma hipersuperfície nula \cite{townsend}.

As hipersuperfícies nulas têm algumas propriedades interessantes. Se um dado vetor tangente a ela deve ser ortogonal ao seu vetor normal, então, ${\bf l\cdot l}=0$ implica que o vetor normal é, ele mesmo, tangente a hipersuperfície nula. Assim,
\begin{equation}
	l^{\mu}=\frac{dx^{\mu}}{d\lambda},
\end{equation}
e as curvas $x^{\mu}(\lambda)$ serão geodésicas.

Geodésicas são curvas onde o vetor tangente a elas se desloca paralelamente a ele mesmo, isto é, um vetor propagado paralelamente a uma geodésica é proporcional ao seu vetor tangente. Com isso, podemos escrever que
\begin{align}
  &l^{\mu}(\nabla_{\mu}l^{\nu})=g(\lambda)l^{\mu},\\
	&l^{\mu}(\partial_{\mu}l^{\nu} + \Gamma^{\nu}_{\rho\mu}l^{\rho}) = g(\lambda)l^{\mu},\\
	&\frac{dx^{\mu}}{d\lambda}\left[\frac{d}{dx^{\mu}}\left(\frac{dx^{\nu}}{d\lambda}\right) + \Gamma^{\nu}_{\rho\mu}\frac{dx^{\rho}}{d\lambda}\right] = g(\lambda)l^{\mu},
\end{align}
resultando na equação da geodésica,
\begin{equation}
	\frac{d^2x^{\mu}}{d\lambda^2} + \Gamma^{\nu}_{\rho\mu}\frac{dx^{\rho}}{d\lambda}\frac{dx^{\mu}}{d\lambda} = g(\lambda)l^{\mu}.
\end{equation}

A função de proporcionalidade $g(\lambda)$ pode ser escolhida de tal forma que $l^{\mu}(\nabla_{\mu}l^{\nu})=0$, sendo $\lambda$ um parâmetro afim.

Note que
\begin{equation}
	\boldsymbol{\xi} \cdot \bold{l} = g_{\mu\nu}l^{\mu}\xi^{\nu}= g_{0\mu}\frac{dx^{\mu}}{\lambda},
\end{equation}
é nulo para $r=2m$. Essa é a definição de um horizonte de Killing, uma superfície nula que tem os vetores de Killing como normais. Daí surge mais uma razão para hipersuperfície em $r=2m$ ser chamada de horizonte de eventos. No horizonte de eventos os vetores de Killing são proporcionais aos vetores normais à hipersuperfície, ou seja,
\begin{equation}
	\xi^{\mu}=h(x)l^{\mu}, \label{f35a}
\end{equation}
para alguma função $h$. Então,
\begin{align}
	&l^{\mu}(\nabla_{\mu}l^{\nu})=0,\\
	&\frac{1}{h}~\xi^{\mu}\nabla_{\mu}\left(\frac{1}{h}~\xi^{\nu}\right)=0, \\
	&\frac{1}{h^2}~\xi^{\mu}\nabla_{\mu}\xi^{\nu} - \frac{1}{h^2}~\xi^{\mu}\xi^{\nu}\partial_{\mu}\ln{\left|h\right|}=0, \\
  &\xi^{\mu}\nabla_{\mu}\xi^{\nu}= \kappa\xi^{\mu},	\label{f35}
\end{align}
onde $\kappa= \xi^{\mu}\partial_{\mu}\ln{\left|h\right|}$ é a chamada de gravidade superficial.

Utilizando (\ref{f35a}) podemos escrever que
\begin{equation}
	\xi_{\mu}\nabla_{\nu}\xi_{\rho}=hl_{\mu}(\nabla_{\nu}h)l_{\rho} + h^2l_{\mu}(\nabla_{\nu}l_{\rho}).
\end{equation}
Note que, do lado direito desta equação, o primeiro termo é simétrico nos índices $\mu$ e $\rho$, enquanto o segundo termo tem simetria nos índices $\nu$ e $\rho$. Com isso, obtemos que sua parte anti-simétrica é nula, ou seja,
\begin{equation}
	\left.\xi_{[\mu}\nabla_{\nu}\xi_{\rho]}\right|_{r=2m}=0. \label{f35b}
\end{equation}
Como $\xi_{\mu}$ é um vetor de Killing, ele satisfaz a equação
\begin{equation}
	 L_{\xi}g_{\mu\nu}=\nabla_{\nu}\xi_{\mu} + \nabla_{\mu}\xi_{\nu}=0,
\end{equation}
onde $L_{\xi}$ é a derivada de Lie em relação ao vetor de Killing. Inserindo este resultado na equação (\ref{f35b}), obtemos
\begin{equation}
	\left[\xi_{\rho}\nabla_{\mu}\xi_{\nu} + \xi_{\mu}\nabla_{\nu}\xi_{\rho} - \xi_{\nu}\nabla_{\mu}\xi_{\rho}\right]_{r=2m}=0.
\end{equation}
Multiplicando ambos os lados da igualdade por $\nabla^{\mu}\xi^{\nu}$, temos que
\begin{equation}
\left[\xi_{\rho}(\nabla_{\mu}\xi_{\nu})(\nabla^{\mu}\xi^{\nu})+2(\nabla^{\mu}\xi^{\nu})\xi_{\mu}(\nabla_{\nu}\xi_{\rho})\right]_{r=2m}=0.
\end{equation}
Mas, utilizando (\ref{f35}), podemos escrever
\begin{align}
\left.(\nabla^{\mu}\xi^{\nu})\xi_{\mu}(\nabla_{\nu}\xi_{\rho})\right|_{r=2m}&=\left.\kappa\xi^{\nu}\nabla_{\mu}\xi_{\rho}\right|_{r=2m}\nonumber \\ &=\left.\kappa^2\xi_{\rho}\right|_{r=2m}.
\end{align}
Portanto, obtemos que
\begin{equation}
	\kappa^2=\left.-\frac{1}{2}\left(\nabla^{\mu}\xi^{\nu}\right)\left(\nabla_{\mu}\xi_{\nu}\right)\right|_{r=2m}. \label{f36}
\end{equation}

A equação (\ref{f36}) é uma forma explícita de se obter a gravidade superficial do horizonte de eventos através dos vetores de Killing. Utilizando a métrica de Schwarzschild, onde $g_{\mu\nu}=diag(g_{tt}, g_{rr}, 1, 1)$ e $\xi^{\mu}=(1, 0, 0, 0)$, podemos chegar a uma expressão da gravidade superficial em função dos elementos da métrica:
\begin{align}
 \kappa^2&=\left.-\frac{1}{2}(g^{\mu\alpha}\nabla_{\alpha}\xi^{\nu})\nabla_{\mu}(g_{\nu\rho}\xi^{\rho})\right|_{r=2m}\nonumber \\
 &=\left.-\frac{1}{2}g^{\mu\alpha}g_{\nu\rho}\Gamma^{\nu}_{t\alpha}\Gamma^{\rho}_{t\mu}\right|_{r=2m}\nonumber \\
&=-\frac{1}{8}\left[g^{tt}g_{rr}\left(-g^{rr}\frac{d}{dr}g_{tt}\right)^2+g^{rr}g_{tt}\left(g^{tt}\frac{d}{dr}g_{tt}\right)^2\right]_{r=2m}.\label{f37}
\end{align}
Como na métrica de Schwarzschild ainda temos
\begin{equation}
	g_{rr}g^{rr}=g_{tt}g^{tt}=1,~~g^{rr}g^{tt}=-1,\label{f38a}
\end{equation}
obtemos
\begin{equation}
	\kappa=\left.\frac{1}{2}\frac{d}{dr}\left(-g_{tt}\right)\right|_{r=2m}, \label{f38}
\end{equation}
uma forma direta e simples para a gravidade superficial de um buraco negro de Schwarzschild. Substituindo a expresão para o termo temporal da métrica concluímos que
\begin{equation}
	\kappa=\frac{1}{4m},\label{f39}
\end{equation}
ou em unidades do sistema internacional
\begin{equation}
	\kappa=\frac{c^3}{4GM}.
\end{equation}

\section{Aceleração no horizonte}
Considere um corpo de massa $m_0$ inicialmente em repouso em $r_0$. Vamos calcular a aceleração desse corpo supondo-o em queda livre. Seu momento é obtido das equações (\ref{f12}) e (\ref{f13})
\begin{equation}
	p^{\mu}=\left(\frac{m_0E}{1-2m/r}, \pm m_0(E^2 - 1 +2m/r)^{1/2}, 0, 0\right),
\end{equation}
onde $E=\sqrt{1-2m/r_0}$ pois $p^1(r_0)=0$. A velocidade desse corpo, calculada em relação a um observador que se move lentamente em $r_0$ é dada por
\begin{equation}
	v=\frac{dist\hat{a}ncia~~pr\acute{o}pria}{tempo~~pr\acute{o}prio}=\frac{1}{1-2m/r}\frac{dr}{dt}.\label{f39a}
\end{equation}
Utilizando a regra da cadeia, é possível escrever a derivada acima como
\begin{equation}
	\frac{dr}{dt}=m_0\frac{dr}{d\tau}\frac{1}{m_0}\frac{d\tau}{dt}=\frac{p^1}{p^0}.
\end{equation}
Substituindo este resultado na equação (\ref{f39a}), concluímos que
\begin{equation}
	v=\left(\frac{2m/r-2m/r_0}{1-2m/r_0}\right)^{1/2}.
\end{equation}
A aceleração medida por este mesmo observador é a variação dessa velocidade pelo tempo próprio, $\tau_p$, tal que $dt/d\tau=(1-2m/r)^{-1/2}$. Segue que
\begin{align}
	a&=\frac{dv}{d\tau_p}=\frac{dv}{dr}\frac{dr}{dt}\frac{dt}{d\tau_p},\nonumber \\
	&=-\frac{m}{r^2}\frac{\sqrt{1-2m/r}}{1-2m/r_0}.
\end{align}
Então, a aceleração para o corpo em repouso em $r_0$ será
\begin{equation}
	a(r_0)=-\frac{m}{r_0^2}(1-2m/r_0)^{-1/2}\label{f40}.
\end{equation}

Agora iremos considerar uma situação hipotética para obter o limite dessa aceleração quando $r\rightarrow2m$. Então, imagine que um observador distante esteja segurando uma partícula de massa unitária por um fio ideal. A energia gasta por ele para suspender essa partícula de um raio $r$ por uma distância $dl$ é $dE_{\infty}=g_{\infty}dl$, onde $g_{\infty}$ é a métrica calculada na posição do observador. A energia da partícula aumenta de $dE_r=g_rdl$, onde $g_r$ é a métrica calculada no ponto onde se encontra a partícula. A conservação da energia implica que
\begin{equation}
	dE_r=\left(1-\frac{2m}{r}\right)^{-1/2}dE_{\infty}~~\Leftrightarrow ~~ g_{00}=g_r\left(1-\frac{2m}{r}\right)^{1/2};
\end{equation}
mas, $g_r$ é justamente a aceleração sentida por um observador em $r$ que foi obtida na equação (\ref{f40}). Portanto, tomando o limite quando a partícula se aproxima do horizonte de eventos, temos que
\begin{equation}
	g_{\infty}=\frac{1}{4m}=\kappa.
\end{equation}
Assim sendo, a força necessária para manter uma partícula de massa unitária sobre o horizonte de eventos de um buraco negro é justamente igual à gravidade superficial do buraco negro. Isto equivale a dizer que a gravidade superficial é a aceleração de uma partícula estática próxima do horizonte medida por um observador no infinito.

\section{Temperatura Hawking}
Depois dos primeiros cálculos realizados por Hawking várias outras formas foram desenvolvidas para se obter a radiação emitida por um buraco negro (veja por exemplo \cite{glauber}). Usaremos um método para investigar os efeitos quânticos nas proximidades do horizonte de eventos conhecido como euclideanização da métrica de Schwarzschild. Ele consiste em uma rotação na coordenada temporal definindo uma coordenada de tempo imaginária,
\begin{equation}
	t=i\tau.
\end{equation}
Aqui, $\tau$ representa o "tempo imaginário"~e não o tempo próprio como de costume.

Esta transformação nos leva à métrica de Schwarzschild euclidiana,
\begin{equation}
	ds_E^2=\left(1-\frac{2m}{r}\right)d\tau^2+\left(1-\frac{2m}{r}\right)^{-1}dr^2+r^2d\Omega^2,\label{f41}
\end{equation}
que também é singular em $r=2m$. Com uma transformação na coordenada radial,
\begin{equation}
	r-2m=\frac{x^2}{8m},\label{f42}
\end{equation}
obtemos que, quando $r\rightarrow2m$ ($x\rightarrow0$),
\begin{align}
&1-\frac{2m}{r}=\frac{(\kappa x)^2}{1+(\kappa x)^2}\approx(\kappa x)^2,\\
&dr^2=(\kappa x)^2dx^2.
\end{align}
Deste modo, teremos
\begin{equation}
	ds_E^2\approx (\kappa x)^2d\tau^2+dx^2+\frac{1}{4\kappa^2}d\Omega^2.\label{f43}
\end{equation}
Este é o chamado espaço-tempo euclidiano de Rindler, pois é obtido através da euclidianização da métrica de Schwarzschild seguido da transformação de Rindler dada pela equação (\ref{f42}). Este sistema de referência é centrado no horizonte de eventos e é simétrico sobre esta origem.

Quando reescrevemos a equação (\ref{f41}), tal que
\begin{equation}
	ds_E^2=dx^2+x^2d(\kappa\tau)^2,\label{f44}
\end{equation}
ela fica idêntica ao espaço euclidiano para coordenadas polares. Porém, para evitarmos a singularidade cônica em $x=0$ precisamos impor uma periodicidade sobre $\tau$ de $2\pi/\kappa$ tal como é para o ângulo polar no espaço euclidiano.

Agora note que, no espaço euclidiano com o tempo imaginário, a integral funcional de um sistema quântico governado por uma hamiltoniana $H$ é
\begin{equation}
	Z=\int[D\phi]e^{-S_E[\phi]},
\end{equation}
onde $S[\phi]$ é a ação e os campos escalares $\phi(\vec{x},t)$ são periódicos na coordenada $t$ com período $\beta\hbar$. Com essas condições a integral funcional pode ser escrita como
\begin{equation}
	Z=tr~e^{-\beta H},
\end{equation}
que é exatamente a função partição do sistema com uma temperatura $T$ dada por $\beta=1/k_BT$, onde $k_B$ é a constante de Boltzman \cite{das}.

Aplicando estes resultados da Teoria Quântica de Campos no caso do espaço euclidiano de Rindler, o que equivale a escrever a periodicidade de $2\pi/\kappa$ na coordenada $\tau$ como sendo $\beta\hbar$, obtemos
\begin{equation}
	T_H=\frac{\kappa\hbar}{2\pi k_B},\label{f45}
\end{equation}
ou, para $\hbar=k_B=1$,
\begin{equation}
	T_H=\frac{\kappa}{2\pi}. \label{f46}
\end{equation}

Esta é a chamada temperatura Hawking, a temperatura de equilíbrio do horizonte de eventos de um buraco negro.

Da lei de Stephan-Boltzmann para a radiação de um corpo negro temos que
\begin{equation}
	\frac{dE}{dt}\cong-\sigma AT_H^4,
\end{equation}
onde $A$ é a área do corpo negro, no nosso caso a área do horizonte de eventos. A constante $\sigma$ é dada por
\begin{equation}
	\sigma=\frac{1}{60}\frac{\pi^2k_B^4}{\hbar^3 c^2}.
\end{equation}
Como
\begin{equation}
	E=Mc^2, ~~~~ A=16\pi m^2,
\end{equation}
utilizando o valor da gravidade superficial no sistema internacional de unidades, temos que
\begin{equation}
	k_BT_H=\frac{\hbar c^3}{8\pi GM}.
\end{equation}
Portanto,
\begin{equation}
	\frac{dM}{dt}\approx -\frac{\hbar c^4}{G^2M^2},
\end{equation}
e, integrando de um tempo inicial onde o buraco negro teria massa $M$ até um tempo final onde toda sua massa tenha evaporado ($M=0$), isso resulta em um tempo de vida
\begin{equation}
	\tau\approx\left(\frac{G^2}{\hbar c^4}\right)M^3. \label{f47}
\end{equation}

Como era esperado, quanto mais massivo for o buraco negro mais tempo ele vive e menor é o valor da sua temperatura Hawking. Se substituirmos os valores das constantes em (\ref{f45}), temos que
\begin{equation}
	T_H \propto \frac{1}{M}\cdot 10^{23}.
\end{equation}
Se um buraco negro tiver sido formado por um colapso gravitacional de uma estrela, sua massa seria da ordem de $10^{30}$ Kg (ordem da massa solar por exemplo). Isso equivaleria a uma temperatura Hawking de $10^{-7}$ K, o que é um valor muito baixo para ser detectado. Portanto, se existir hoje no universo buracos negros de natureza estelar, eles não seriam úteis para detectar a radiação Hawking.

Poderíamos então buscar casos onde a evaporação estaria quase no fim, com a massa do buraco negro sendo muito pequena. Isso possivelmente causaria uma explosão de radiação extremamente intensa. Mas como o tempo necessário para que toda a massa se evapore é proporcional a $M^3$, novamente vemos que buracos negros estelares seriam inapropriados pois seu tempo médio de vida seria da ordem de $10^{73}$ anos, o que é muito maior que a própria idade do universo.

No entanto, buracos negros com massas bem menores (da ordem de $10^{12}$ Kg), poderiam se formar a partir da extrema densidade de matéria presente na expansão inicial do universo. Conhecidos como \textit{buracos negros primordiais}, estes sim estariam hoje em seus estágios finais de evaporação. O problema aqui decorre do fato de se encontrar realmente estes buracos negros primordiais, visto que seu tamanho seria muito pequeno e com pouca influência gravitacional.

Devemos ressaltar aqui que os cálculos originais de Hawking não levam em consideração a influência que a propagação dos fótons têm sobre as equações da mecânica quântica. Isto é válido enquanto a massa do buraco negro é maior do que a taxa de energia emitida pela radiação. Porém, nos estágios finais dessa evaporação, quando a massa é da ordem da energia emitida, seria necessário uma teoria quântica da gravitação consistente para realizar tais cálculos.

\section{Teorema da Área}
Para um buraco negro de Schwarzschild, a área do horizonte de eventos é calculada a partir do elemento de área de uma esfera de raio $r_h$:
\begin{align}
	dA&=8\pi r_hdr_h, \\
	dA&=32\pi mdm,\\
	dA&=8\pi\frac{G}{c\kappa}dM,
\end{align}
ou ainda,
\begin{equation}
	dM=\frac{\kappa c}{8\pi G}dA.
\end{equation}
Substituindo a gravidade superficial pela temperatura Hawking,
\begin{equation}
	dM=T_H\frac{k_Bc}{4\hbar G}dA,\label{f49}
\end{equation}
podemos definir
\begin{equation}
	S=\frac{k_B}{4\hbar G}A\label{f50}
\end{equation}
e, utilizando $E=Mc^2$, obtemos
\begin{equation}
	dE=T_HdS.\label{f51}
\end{equation}
Portanto, a equação (\ref{f49}) representa a primeira lei da termodinâmica para um buraco negro de Schwarzschild, com a entropia dada pela equação (\ref{f50}).

Note que, enquanto o buraco negro emite essa radiação sua massa vai diminuindo, fato que ficou explícito quando obtemos uma taxa negativa para a variação da massa com o tempo. Isso diretamente implica na redução da área do horizonte de eventos à medida que o buraco negro vai evaporando e no aumento da taxa de energia irradiada pelo buraco. Até aí os resultados da teoria de Hawking são satisfatórios. Porém, quando se faz a analogia da mecânica do buraco negro com as leis da termodinâmica, impomos que a entropia do sistema é diretamente proporcional a área do horizonte de eventos. Isso implica em uma quebra imediata da segunda lei da termodinâmica, que diz que a variação da entropia de um sistema nunca deve ser menor que zero.

Este impasse na formulação de uma teoria termodinâmica para buracos negros foi resolvido também por Hawking \cite{hawking2} quando ele notou que a entropia do buraco negro diminui com sua evaporação mas a entropia do universo externo ao horizonte de eventos aumenta com a radiação que chega até ele, e esse acréscimo é maior que a variação da entropia do buraco negro. Assim, a segunda lei da termodinâmica é reformulada tomando
\begin{equation}
	S=S_{BN}+S_{ext},\label{f52}
\end{equation}
onde $S_{BN}$ é a entropia do buraco negro dado por (\ref{f50}) e $S_{ext}$ é a entropia do universo exterior ao buraco negro. Com isso, a variação da entropia será sempre maior que zero, respeitando a segunda lei da termodinâmica.


\chapter{Análogo Acústico a um Buraco Negro}

Os sistemas análogos para a relatividade geral começaram a ser desenvolvidos no início da década de oitenta com o físico canadense Willian G. Unruh \cite{unruh}. Unruh demonstrou que ondas sonoras se propagando em um fluido em movimento é fenômeno semelhante ao da luz quando viaja em um espaço-tempo curvo. As ondas sonoras também sofrerão uma variação de frequência à medida que se propagam, contra ou a favor, da direção de escoamento do fluido. Em especial, se o fluido atinge velocidades supersônicas, em um determinado ponto se formará uma barreira sonora tal como o horizonte de eventos de um buraco negro gravitacional. Assim, as ondas sonoras que estiverem dentro da região onde o fluido se movimenta mais rápido que a velocidade do som, nunca conseguirão escapar, definindo assim o \textit{buraco negro acústico}.

Mais interessante ainda é que a propagação de ondas sonoras na vizinhança desse buraco negro acústico pode ser descrita pelo ferramental geométrico da relatividade geral através de uma métrica efetiva. Com isso, este modelo análogo aos buracos negros gravitacionais, se torna extremamente útil no estudo dos fenômenos existentes na física de buracos negros. Entre eles a radiação Hawking, principal alvo de estudo dessa dissertação.

Neste capítulo iremos demonstrar como essa analogia surge nas equações matemáticas da mecânica dos fluidos e analisar o que ela nos revela de novo no campo da física de buracos negros e, consequentemente, no estudo do efeito Hawking.

\section{A métrica acústica}
Com o intúito de obter uma análise mais completa possível, iremos nos concentrar em um modelo acústico análogo ao buraco negro gravitacional de Schwarzschild. Para este fim, devem ser impostas algumas considerações iniciais sobre o nosso sistema acústico. Trabalharemos com um fluido em movimento, tal que, em um certo ponto, ele atinja e supere a velocidade do som. Não iremos nos deter inicialmente a qualquer configuração especial que realize este feito, basta que ele ocorra. Sob estas condições, o fluido deve ser irrotacional para que a não haja vórtices e assim a velocidade do fluido possa ser descrita inteiramente por um campo escalar. Ele deve ser barotrópico também, ou seja, sua pressão será função somente de sua própria densidade. Isso garante que um fluido inicialmente irrotacional permaneça irrotacional. Por fim, para simplificar, consideramos o fluido não viscoso. O efeito causado quando se leva em consideração a viscosidade do fluido pode ser estudado em \cite{visser}.

A dinâmica do fluido é descrito por três equações \cite{landau}:
\begin{itemize}
	\item A equação da continuidade
	 \begin{eqnarray}
    \frac{\partial\rho}{\partial t}+\boldsymbol{\nabla}\cdot(\rho\bf{v}) = 0; \label{f301}
   \end{eqnarray}
  \item a equação de Euler
   \begin{eqnarray}
	  \rho\left[\frac{\partial \bf{v}}{\partial t} +\left(\bf{v}\cdot\boldsymbol{\nabla}\right)\bf{v}\right]   				=-\boldsymbol{\nabla}p-\rho\boldsymbol{\nabla}\Phi;\label{f302}
   \end{eqnarray}
  \item e alguma equação barotrópica
   \begin{eqnarray}
    p=p(\rho),\label{f303}
   \end{eqnarray} 
\end{itemize}
com $\rho$ sendo a densidade do fluido, $\bf v$ a velocidade, $p$ a pressão e $\Phi$ representa tanto o potencial gravitacional newtoniano quanto um potencial externo qualquer.

A propagação do som é descrita matematicamente como uma perturbação linear das variáveis dinâmicas do meio. Desta forma, como na radiação Hawking, que é baseada na separação dos pares virtuais de fótons provenientes das flutuações quânticas do estado de vácuo, espera-se que as perturbações do fluido nas proximidades do horizonte de eventos acústico possa originar uma emissão de ondas sonoras tal como o efeito Hawking \cite{visser2}.

Então, vamos linearizar estas equações com perturbações em primeira ordem em torno de alguma solução exata ($\rho_0, p_0, \psi_0$) das equações (\ref{f301}), (\ref{f302}) e (\ref{f303}). Como escolhemos que a velocidade do fluido seja irrotacional, ou seja, $\boldsymbol{\nabla}\times {\bf v}={\bf0}$, podemos definir esta velocidade como
\begin{equation}
	{\bf v}= -\boldsymbol{\nabla}\psi,
\end{equation}
onde $\psi$ é um campo escalar chamado potencial velocidade. Usando este potencial velocidade, as quantidades perturbadas serão tais que,
\begin{eqnarray}
	\rho=\rho_0 + \epsilon\rho_1+ \mathcal{O}(\epsilon^2); \nonumber \\
	p=p_0 + \epsilon p_1+ \mathcal{O}(\epsilon^2);\label{f305a} \\
	\psi = \psi_0 + \epsilon\psi_1+\mathcal{O}(\epsilon^2),\nonumber
\end{eqnarray}
onde os $\mathcal{O}(\epsilon^2)$ representam os termos de segunda ordem em diante, que serão negligenciados nos nossos cálculos. Note que estamos mantendo o potencial $\Phi$ fixo, isto é, estamos impondo que não haja reações de fundo. Estas reações seriam a deformação do potencial $\Phi$ causada pela propagação das ondas $\psi_1$.

Reescrevemos as equações (\ref{f301}~-~\ref{f303}) utilizando as perturbações (\ref{f305a}), obtendo assim,
\begin{eqnarray}
\frac{\partial\rho_1}{\partial t}+\boldsymbol{\nabla}\cdot\left(\rho_1\boldsymbol{\nabla}\psi_0+\rho_0\boldsymbol{\nabla}\psi_1\right)=0,\label{f304}\\
	\rho_0\left(\frac{\partial\psi_1}{\partial t}+\boldsymbol{\nabla}\psi_0\cdot\boldsymbol{\nabla}\psi_1\right)=p_1,\label{f305}\\
	p_1=c^{2}_{s}\rho_1.\label{f306}
\end{eqnarray}

Para a última equação acima, utilizamos a definição padrão da velocidade do som, $c_s^2={\partial p}/{\partial\rho}$, e a tomamos como uma constante. A combinação dessas três equações resulta em uma equação diferencial de segunda ordem para $\psi_1$
\begin{equation}
-\frac{\partial}{\partial t}\left[\frac{\rho_0}{{c_s}^2}\left(\frac{\partial\psi_1}{\partial t} + \bf{v_0}\cdot\boldsymbol{\nabla}\psi_1\right)\right] + \boldsymbol{\nabla}\cdot\left[\rho_0\boldsymbol{\nabla}\psi_1 - \frac{\rho_0}{{c_s}^2}\bf{v_0}\left(\frac{\partial\psi_1}{\partial t} + \bf{v_0}\cdot\boldsymbol{\nabla}\psi_1\right)\right]=0.\label{f307}
\end{equation}

Esta é a equação de onda que descreve a propagação do potencial velocidade. Uma vez que tenhamos determinado o valor de $\psi_1$ podemos utilizar as equações (\ref{f304}) e (\ref{f305}) para determinarmos $p_1$ e $\rho_1$. Portanto, a equação de onda acima descreve completamente a propagação das perturbações acústicas.

Agora, nosso objetivo é descrever a propagação dessas perturbações acústicas tal como o formalismo da relatividade geral. Para tal, introduzimos uma matriz 4 x 4 \cite{visser}, definida como
\begin{eqnarray}
f^{\mu\nu}\equiv \frac{\rho_0}{c^2_s}\left[\begin{array}{ccc}
-1 & \vdots & -v^{j}_{0}\\
\cdots\cdots\cdots\cdots\cdots & \cdot & \cdots\cdots\cdots \\
-v^{i}_{0} & \vdots & (c^{2}_{s}\delta^{ij}- v^{i}_{0} v^{j}_{0})\end{array}\right]. \label{f308}
\end{eqnarray}

Assim, podemos reescrever a equação (\ref{f307}) como
\begin{eqnarray}
\partial_{\mu}\left(f^{\mu\nu}\partial_{\nu}\psi_1\right)=0, \label{f309}
\end{eqnarray}

mas o d'Alembertiano no espaço-tempo curvo \cite{weinberg} é definido como
\begin{eqnarray}
\Delta\psi\equiv\frac{1}{\sqrt{-g}}\partial_{\mu}\left(\sqrt-g g^{\mu\nu}\partial_{\nu}\psi\right), \label{f310}
\end{eqnarray}
onde $g^{\mu\nu}$ é a inversa da métrica do espaço-tempo que se está calculando o d'Alembertiano e $g$ é o seu determinante. Portanto, impondo a condição $f^{\mu\nu}=\sqrt{-g}g^{\mu\nu}$, obtemos nova matriz que representa uma métrica para o nosso sistema, que chamaremos de métrica acústica,
\begin{eqnarray}
g_{\mu\nu}^{ac\acute{u}stica}=\frac{\rho_0}{c_s}\left[\begin{array}{ccc}
-({c_s}^2-{v_0}^2) & \vdots & -v_0^j\\
\cdots\cdots\cdots\cdots\cdots & \cdot & \cdots\cdots\cdots \\
-v_0^i & \vdots & \delta_{ij}\end{array}\right]. \label{f311}
\end{eqnarray}
Com a métrica em mãos, podemos escrever o elemento de linha acústico
\begin{eqnarray}
ds^2=\frac{\rho_0}{c_s}\left[-c^2_sdt^2+\delta_{ij}(dx^i-v_0^idt)( dx^j-v_0^jdt)\right]. \label{f312}
\end{eqnarray}

Observe que a assinatura da métrica acústica, $(-, +, +, +)$, é exatamente como deve ser em uma geometria Lorentziana. É muito importante entender o que a métrica acústica representa no nosso sistema. Na verdade, o nosso sistema físico está situado em uma variedade descrita pela métrica usual de Minkowski, com a topologia básica do espaço ${\Re}^4$. Contudo, as perturbações acústicas (o campo escalar $\psi_1$) se comportam como se "vivessem" ~em uma variedade cuja métrica é dada pela equação (\ref{f312}). Tanto que essa métrica acústica herda a topologia e a geometria do espaço-tempo de Minkowski. Devemos enfatizar também que a métrica na relatividade geral está relacionada a distribuição de matéria no espaço-tempo, enquanto a métrica acústica está relacionada a distribuição de matéria de uma forma algébrica \cite{visser, visser3}. Mais corretamente, ela descreve uma espécie de distribuição de velocidade do fluido, que age sobre o som tal como a massa sobre a luz.

\subsection{Ergo-regiões e horizonte de eventos acústico}
Considere os vetores de Killing $\xi^{\mu}=(1, 0, 0, 0)$, tal que
\begin{equation}
	g_{\mu\nu}\xi^{\mu}\xi^{\nu}=g_{tt}=-(c^2-v^2).
\end{equation}
Esta expressão será positiva para $\left\|\bf{v}\right\|>0$. Portanto, qualquer região supersônica será chamada de ergo-região, o equivalente a ergo-esfera existente na geometria de Kerr para buracos negros com rotação. Na relatividade geral essa região é onde o espaço "se move"~mais rápido que a velocidade da luz, sendo impossível permanecer em repouso lá. Na analogia acústica, tentar ficar parado em uma ergo-região implicaria em uma turbulência sônica.

Na métrica acústica podemos ver explicitamente a existência de um horizonte de eventos, ocorrendo quando a velocidade do fluido se iguala à velocidade local do som ($v_0 = c_s$). Um horizonte de eventos acústico é definido pela superfície que delimita a região da qual geodésicas nulas (os fônons) não podem escapar. Em toda geometria estacionária, como a de Schwarzschild, o horizonte de eventos é a borda da ergo-região. No nosso sistema acústico ocorre da mesma forma pois o horizonte separa as regiões subsônica e supersônica.

Uma dos modelos mais simples para estudar os horizontes de eventos acústicos é de um fluido em movimento onde suas velocidades transversais são muito menores que a velocidade no eixo $x$, podendo assim ser negligenciadas. Com isso podemos obter o elemento de linha acústico para um fluido em movimento unidimensional como
\begin{equation}
	 ds^2= \frac{\rho_0}{c_s}\left[-(c_s^2-v_0^2)dt^2 - 2v_0dtdx + dx^2 + dy^2 + dz^2\right], \label{f313}
\end{equation}
onde $v_0$ é o módulo da velocidade do fluido, que é um vetor no eixo $x$.

\section{Analogia à métrica de Schwarzschild}
Obtida a métrica acústica, o próximo passo é verificar se realmente podemos imitar a métrica de Schwarzschild e de que maneira essa analogia deve ser feita. Para tal, começamos com uma transformação na coordenadas de Schwarzschild para obter uma representação que se adapta mais a equação (\ref{f312}). Partimos da seguinte transformação na coordenada temporal \cite{kraus},
\begin{equation}
	 t=t_s + \left[2\sqrt{2GMr}-2GMln\left(\frac{\sqrt{2GMr}+2GM}{\sqrt{2GMr}-2GM} \right)\right], \label{f314}
\end{equation}
onde $t_s$ representa a coordenada temporal de Schwarzschild. Dessa transformação obtemos
\begin{equation}
	dt_s = dt - \frac{\sqrt{2GM/r}}{(1 - 2GM/r)}dr. \label{f315}
\end{equation}
Substituindo na equação (\ref{f5}), com $c=1$, temos que
\begin{equation}
	ds^2= -\left(1-\frac{2GM}{r}\right)dt^2 + 2\sqrt{\frac{2GM}{r}}dtdr + dr^2 + r^2d\Omega^2, \label{f317}
\end{equation}
onde $d\Omega$ é a parte angular do elemento de linha em coordenadas esféricas. Esta nova representação da métrica de Schwarzschild é conhecida como as coordenadas de Painlevè-Gullstrand, \cite{PG}, e ela cobre toda região assintótica, o horizonte futuro e a singularidade do buraco negro. Essa forma também é não singular no horizonte e transfere toda curvatura do espaço-tempo de Schwarzschild para a coordenada temporal, ou seja, para $t=cte$ a parte espacial é plana.

Comparando o elemento de linha (\ref{f317}) com o elemento de linha acústico (\ref{f313}) notamos que, a menos de um fator conforme, eles são idênticos (visto que o elemento de linha acústico também pode ser escrito em coordenadas esféricas). Então, a analogia aconteceria de forma exata se escolhêssemos um fluido cuja densidade fosse independente da posição, tomássemos a velocidade do som também como constante e definíssemos o perfil da velocidade do fluido como sendo
\begin{equation}
	v_0=\sqrt{2GM/r}. \label{f318}
\end{equation}
Porém, esta configuração do nosso sistema acústico não estaria de acordo com a equação da continuidade (\ref{f301}) que deve ser satisfeita, pois a usamos para deduzir a própria métrica acústica. Isso nos leva a concluir que uma métrica acústica conforme com a métrica do sistema gravitacional é o mais próximo que podemos chegar com essa analogia.

Contudo, isso é extremamente satisfatório para análise da radiação Hawking, visto que as grandezas envolvidas nessa teoria (gravidade superficial em especial) são invariantes perante transformações conformes da métrica \cite{jacobsen}. Este modelo então nos fornece um sistema munido de um horizonte de eventos que, assim como na relatividade geral, emite fônons segundo o argumento de Hawking.

\section{Gravidade superficial e Temperatura Hawking}
A radiação Hawking é caracterizada exclusivamente pela sua temperatura que por sua vez é obtida diretamente da gravidade superficial do buraco negro. Portanto, se definirmos uma gravidade superficial para o sistema acústico, conseguimos obter uma equação para a temperatura Hawking equivalente. Para isso, iremos inicialmente transformar a métrica acústica (\ref{f313}) para a forma tradicional de Schwarzschild usando para a nova coordenada temporal
\begin{equation}
	\tau=t+\left[\frac{2vx}{c_s^2}-\frac{v^2x}{c_s^3}\ln\left({\frac{c_s+v}{c_s-v}}\right)\right],
\end{equation}
tal que
\begin{equation}
	d\tau=dt+\frac{v}{c_s^2-v^2}dx. \label{f319}
\end{equation}
Esta nova coordenada temporal, $\tau$, é sugerida a partir da transformação (\ref{f314}) usando $v\propto\sqrt{2GM/r}$ e fazendo os ajustes de dimensão com $c_s$. Substituindo (\ref{f319}) no elemento de linha acústico obtemos
\begin{equation}
	ds^2=\frac{\rho_0}{c_s}\left[-(c_s^2-v^2)d\tau^2+\frac{c_s^2}{c_s^2-v^2}dx^2 +dy^2 + dz^2\right].\label{f320}
\end{equation}
Sob este novo sistemas de coordenadas podemos ver claramente que a geometria acústica é estática, isso devido à ausência dos termos cruzados de tempo e espaço. Então, os vetores de Killing são normais ao horizonte de eventos, tal como foi demonstrado no capítulo anterior. Podemos utilizar a equação (\ref{f37}) para calcular a gravidade superficial acústica, substituindo as componentes da métrica de Schwarzschild pelas componentes da métrica acústica (\ref{f320}), obtendo que
\begin{equation}
	g_{xx}g^{xx}=g_{tt}g^{tt}=1,~~g^{xx}g^{tt}=-\frac{1}{\rho_0^2}.
\end{equation}
Assim,
\begin{equation}
\kappa=\left.\frac{1}{2\rho_0}\frac{d}{dx}\left(-g_{tt}\right)\right|_{v_o=c_s}=\frac{1}{2\rho_0}\frac{d}{dx}\left[\frac{\rho_0}{c_s}(c^2_s-v_0^2)\right]_{v_o=c_s}. \label{f321a}
\end{equation}
Note que a gravidade superficial deve ser calculada quando $v_0=c_s$ que é a definição do horizonte de eventos acústico. Usando a propriedade da derivada do produto de funções, escrevemos
\begin{eqnarray}
	\frac{d}{dx}\left[\frac{\rho_0}{c_s}(c_s^2-v_0^2)\right]_{v_0=c_s}&=&\left[\frac{\rho_0}{c_s}\frac{d}{dx}(c_s^2-v_0^2)     +(c_s^2-v_0^2)\frac{d}{dx}\frac{\rho_0}{c_s}\right]_{v_0=c_s},\nonumber \\
	&=&\left[\frac{\rho_0}{c_s}\frac{d}{dx}(c_s^2-v_0^2)\right]_{v_0=c_s}.\label{f321b}
\end{eqnarray}
Por fim, substituindo (\ref{f321b}) em (\ref{f321a}), obtemos
\begin{equation}
	\kappa=\left.\frac{1}{2c_s}\frac{d}{dx}(c_s^2-v_0^2)\right|_{v_o=c_s}.\label{f321}
\end{equation}
Com isso, utilizando a equação (\ref{f45}) obtemos a temperatura Hawking para o análogo acústico,
\begin{equation}
	T=\left.\frac{\hbar}{4\pi c_sk_B}\frac{d}{dx}(c_s^2-v_0^2)\right|_{v_o=c_s}. \label{f322}
\end{equation}

Dessa forma, o horizonte acústico emite uma radiação Hawking acústica, na forma de um banho térmico de fônons, que é caracterizada pela temperatura acima.

É importante observar que, nossa métrica acústica depende da densidade do fluido, da velocidade do fluxo e da velocidade local do som. Ela é governada pelas equações de movimento do fluido somente, não utilizamos em momento algum, as equações de Einstein da relatividade geral. Ou seja, nosso modelo só pode ser usado para uma analogia com a cinemática do sistema gravitacional. Sendo os aspectos dinâmicos não inclusos neste modelo, ficamos impossibilitados de reproduzir a termodinâmica de um buraco negro. Este resultado não é restrito ao nosso modelo acústico, vários outros modelos que se propõem a imitar a gravidade não estão sujeitos à dinâmica do relatividade geral, veja por exemplo \cite{visser5}.

No entanto, vamos ver no próximo capítulo que, para um buraco negro bidimensional, podemos sim obter uma definição de massa para o buraco negro acústico e assim produzir uma analogia às leis da termodinâmica do sistema gravitacional.


\chapter{Analogia à dinâmica de um buraco negro}
Nas sessões que seguem este capítulo iremos demonstrar que, a partir de um buraco negro bidimensional, podemos obter uma analogia direta entre a dinâmica do sistema gravitacional e a dinâmica do fluido. Veremos que reduzindo um buraco negro 4D esfericamente simétrico a uma descrição 2D, sua dinâmica equivale a uma dinâmica de fluido com um certo vínculo, que nos permitirá definir a primeira lei da termodinâmica para um buraco negro acústico.

\section{Métrica acústica para um buraco negro 2D}
Um sistema gravitacional quadri-dimensional esfericamente simétrico, tal como o espaço-tempo de Schwarzschild, pode ser representado por um sistema bidimensional com um campo dilatônico caracterizado por um potencial $V$, cuja ação é dada por \cite{grumiller, martinez}
\begin{equation}
	S=\frac{1}{2}\int d^2x\sqrt{-g}\left(\phi R+\lambda^2V(\phi)\right),\label{401}
\end{equation}
onde $\phi$ é um campo escalar, o próprio dílaton, e $\lambda$ é um parâmetro com unidade de [comprimento]$^{-1}$. Aqui estamos usando $c=\hbar=k_B=1$. Esta ação admite soluções tipo Schwarzschild que toma a forma
\begin{equation}
	ds^2=-\left(J(\phi)-\frac{2M}{\lambda}\right)dt_s^2+\left(J(\phi)-\frac{2M}{\lambda}\right)^{-1}dr^2, \label{402}
\end{equation}
onde $\phi=\lambda r$, $M$ é a massa do buraco negro e $J=\int Vd\phi$. O horizonte de eventos ocorre quando $J(\phi_h)=2M/\lambda$ e assim $\phi_h=\lambda r_h$. Agora precisamos encontrar uma métrica acústica análoga a equação (\ref{402}).

Do capítulo anterior temos que, a métrica acústica para um fluido ideal de fluxo unidimensional, é escrita como
\begin{equation}
	ds^2=\frac{\overline{\rho_0}}{c_s}\left[-(c_s^2-v_0^2)dt^2-2v_0dxdt+dx^2\right],\label{403}
\end{equation}
e a velocidade local do som é dada por $c^2_s=dP/d\overline{\rho_0}$. Usamos $\overline{\rho_0}$ para identificar a densidade usual, pois iremos trabalhar com uma densidade adimensional $\rho_0=\lambda^{-4}\overline{\rho_0}$. Da mesma maneira como foi feito anteriormente, devemos fazer uma transformação de Painlevè-Gullstrand na métrica da equação (\ref{402}) para fins de comparação com a métrica acústica. Usando a equação (\ref{f318}), que define um perfil de velocidade para o qual o sistema acústico é conforme ao sistema gravitacional, reescrevemos a transformação (\ref{f315}) como sendo
\begin{equation}
	dt_s=dt+\frac{v_0}{c_s^2-v_0^2}dx. \label{404}
\end{equation}
Substituindo isso na métrica (\ref{402}), teremos
\begin{eqnarray}
	ds^2=-\left(J-\frac{2M}{\lambda}\right)dt^2 &-& \left(J-\frac{2M}{\lambda}\right)\frac{2v_0}{c_s^2-v_0^2}dtdx \nonumber \\   
	\nonumber \\
	&-& \left(J-\frac{2M}{\lambda}\right)\frac{v_0^2}{(c_s^2-v_0^2)}dx^2 + \left(J-\frac{2M}{\lambda}\right)^{-1}dr^2.
\end{eqnarray}
Comparando a componente tempo-tempo dessa métrica com a métrica acústica usual escolhemos que
\begin{equation}
	J=\frac{2M}{\lambda}+\frac{\rho_0}{c_s}(c_s^2-v_0^2), \label{405}
\end{equation}
e então obtemos
\begin{equation}
	ds^2=\frac{\rho_0}{c_s}\left[-(c_s^2-v_0^2)dt^2 - 2v_0dtdx -\frac{v_0^2}{c_s^2-v_0^2}dx^2\right] + \frac{c_s^2}{\rho_0^2(c_s^2-v_0^2)}dr^2.
\end{equation}
Por fim, façamos uma transformação na coordenada radial do sistema gravitacional para a coordenada cartesiana do sistema acústico,
\begin{equation}
	dr=\rho_0dx,\label{406}
\end{equation}
e então
\begin{eqnarray}
	ds^2 &=& \frac{\rho_0}{c_s}\left[-(c_s^2-v_0^2)dt^2 - 2v_0dtdx -\frac{v_0^2}{(c_s^2-v_0^2)}dx^2 + \frac{c_s^2}{(c_s^2-v_0^2)}dx^2\right]\nonumber \\
	\nonumber \\ 
	&&~~~~~ ds^2 = \frac{\rho_0}{c_s}\left[-(c_s^2-v_0^2)dt^2 - 2v_0dtdx + dx^2\right].\label{407}
\end{eqnarray}
Assim sendo, vemos que, com as transformações dadas por (\ref{404}), (\ref{405}) e (\ref{406}), a correspondência entre os dois modelos é exata. Isso difere do modelo análogo para um buraco negro 4D, como visto no capítulo anterior, onde a equivalência entre eles só é válida perante transformações conforme da métrica.

No entanto, este modelo é limitado a imitar somente a cinemática de um buraco negro 2D. Para que possamos estender a analogia até um nível dinâmico é preciso comparar a dinâmica gravitacional com a dinâmica do fluido, que determina $\rho_0, v_0$ e $c_s$.

\section{Analogia à dinâmica gravitacional}
A principal diferença entre os dois sistemas físicos tratados nesta dissertação é que o sistema gravitacional deve ser covariante sobre transformações gerais da métrica. A relatividade geral é uma teoria independente do ''background'', ou seja, não é necessário a existência de uma métrica específica para deduzirmos as equações de campo provenientes da ação \cite{alex, ashtekar}. Já no sistema acústico a métrica surge de perturbações nas equações de movimento, sendo então já previamente fixada.

Portanto, para construir uma dinâmica acústica análoga, devemos fixar uma métrica para a descrição do nosso sistema gravitacional dilatônico 2D. Porém, a gravidade dilatônica 2D é puramente topológica, então não temos uma propagação física dos graus de liberdade gravitacionais. Isto deixa em aberto a possibilidade de que, uma vez que tenhamos fixado a métrica, a dinâmica dos graus de liberdade gravitacional possa ser imitada pela dinâmica do fluido.

As equações de campo para o modelo dado em (\ref{401}) são
\begin{align}
	R=-\lambda\frac{dV}{d\phi},\\
	\nabla_{\mu}\nabla_{\nu}\phi-\frac{1}{2}g_{\mu\nu}\lambda^2V=0. \label{408}
\end{align}
Nós fixamos a métrica escolhendo o calibre de Schwarzschild
\begin{equation}
	ds^2=-X(r)dt_s^2 + X(r)^{-1}dr^2, \label{409}
\end{equation}
que é estática, justamente como queremos para que seja compatível com o modelo acústico já apresentado aqui. Utilizando as equações (\ref{408}) e (\ref{409}) obtemos que
\begin{equation}
	\frac{d\phi}{dr}=\lambda,~~~~\phi=\lambda r, \label{410}
\end{equation}
e
\begin{equation}
	\frac{dX}{dr}=\lambda V, \label{411}
\end{equation}
de onde pode-se verificar que o buraco negro (\ref{402}) é solução dessa última expressão.

Isso nos mostra que, quando fixamos o calibre, o sistema gravitacional fica com somente um grau de liberdade, a função $X(r)$ que parametriza a métrica, e uma única equação. O campo escalar $\phi$ tem como vínculo ser diretamente proporcional a $r$, portanto ele representa apenas uma coordenada tipo espaço.

Agora precisamos obter as equações da dinâmica dos fluidos, que são as equações da continuidade e de Euler, respectivamente representadas pelas equações (\ref{f301}) e (\ref{f302}). Mas faremos um tratamento unidimensional. Assim, as equações anteriores se resumem a
\begin{equation}
	\bar{\rho_0}v_0\frac{dv_0}{dx}+\frac{dp}{dx}+\bar{\rho_0}\frac{d\psi}{dx}=0, \label{412}
\end{equation}
e
\begin{equation}
	\bar{\rho_0}(x)v_0(x)A(x)=D,\label{413}
\end{equation}
onde $A$ denota a área da secção transversal por onde este fluido está passando e $D$ é uma constante. Vamos considerar dois casos diferentes para solucionar estas equações:
\begin{description}
	\item[a] - O potencial externo $\psi$ é não homogêneo, sendo dependente da posição, e $A$ é uma constante;
	\item[b] - O potencial externo é nulo e $A$ é uma função dependente da posição $x$.
\end{description}

\subsection*{Caso - a}
Vamos reescrever as equações (\ref{412}) e (\ref{413}) a partir da mudança da coordenada $x$ pela coordenada radial do sistema gravitacional tal como foi feito antes usando (\ref{406}). Assim, usando também a relação entre $\bar{\rho_0}$ e $\rho_0$, e considerando a condição barotrópica que implica que a pressão só dependa da densidade do fluido, temos que
\begin{eqnarray}
	\rho_0v_0\frac{dv_0}{dr} + \frac{dp}{d\bar{\rho_0}}\frac{d\rho_0}{dr} + \rho_0\frac{d\psi}{dr}=0.
\end{eqnarray}
Multiplicando tudo por $2/c_s$ e usando a definição de velocidade do som onde $c_s^2=dp/d\bar{\rho_0}$, temos que
\begin{equation}
	2~\frac{\rho_0v_0}{c_s}\frac{dv_0}{dr} + 2c_s~\frac{d\rho_0}{dr} + 2~\frac{\rho_0}{c_s}\frac{d\psi}{dr}=0.
\end{equation}
Definindo então as novas variáveis
\begin{eqnarray}
	Z&=&\frac{\rho_0}{c_s}(c_s^2-v_0^2),\nonumber \\
	Y&=&\rho_0c_s, \label{414} \\
	F&=&\ln\left(\frac{c_s}{\rho_0}\right),\nonumber
\end{eqnarray}
obtemos
\begin{eqnarray}
	\frac{dZ}{dr}&=&\frac{1}{c_s}\frac{d\rho_0}{dr}(c_s^2-v_0^2) - 2\frac{\rho_0v_0}{c_s}\frac{dv_0}{dr},\nonumber \\
	\nonumber \\
	\frac{dY}{dr}&=&c_s\frac{d\rho_0}{dr},\nonumber \\
	\\
	\frac{dF}{dr}&=&-\frac{1}{\rho_0}\frac{d\rho_0}{dr}, \nonumber \\
	\nonumber \\
	\frac{\rho_0}{c_s}&=& e^{-F}, \nonumber
\end{eqnarray}
para reescrever a equação de Euler como
\begin{equation}
	\frac{dZ}{dr}=2\frac{dY}{dr} - Z\frac{dF}{dr} + 2e^{-F}~\frac{d\psi}{dr}. \label{415}
\end{equation}

Na equação (\ref{413}) escrevemos que
\begin{eqnarray}
	\frac{d}{dr}\left(\frac{\rho_0}{\lambda^4}v_0A\right)=0\nonumber \\
	\nonumber \\
	\frac{d}{dr}(\rho_0v_0)=0,
\end{eqnarray}
mas como
\begin{equation}
	\rho_0v_0^2=\rho_0c_s^2-c_sX ~~~~\Rightarrow~~~~ \rho_0v_0=\sqrt{Y(Y-Z)},
\end{equation}
logo
\begin{equation}
	\frac{d}{dr}\left[Y(Y-Z)\right]=0. \label{416}
\end{equation}

Com as equações (\ref{416}) e (\ref{415}) representando a dinâmica do fluido vemos que, como a analogia entre as métricas é feita tal que $Z=X$ (componentes temporal de cada métrica), a equivalência entre a dinâmica gravitacional e as equações de Euler e da continuidade pode ser estabelecida impondo um vínculo de acordo com a equação (\ref{411}). Ou seja
\begin{equation}
	2\frac{dY}{dr} - Z\frac{dF}{dr} + 2e^{-F}~\frac{d\psi}{dr}=\lambda V. \label{417}
\end{equation}

Isso mostra que, se o potencial externo for dado, a dinâmica do fluido é caracterizada por três funções desconhecidas, $\rho_0, v_0, c_s$ e duas equações. Temos então a liberdade de estabelecer um vínculo como uma terceira equação para o sistema. Na verdade a equação (\ref{417}) fornece uma equação de estado para o fluido para um dado $\psi$, e assim esta equação de estado pode ser vista como uma consequência da dinâmica gravitacional. Por outro lado, se tivermos uma equação de estado previamente definida para o fluido, este vínculo definirá o potencial externo.

A massa para o buraco negro dilatônico em 2D \cite{cadoni, mann},
\begin{equation}
	M=\frac{1}{2\lambda}\left(\lambda^2\int{V(\phi)d\phi}-(\nabla\phi)^2\right),\label{418}
\end{equation}
permanece constante quando calculada no horizonte de eventos. Seguindo esta formulação, podemos obter uma expressão equivalente a massa do buraco negro de Schwarzschild através das equações (\ref{416}) e (\ref{415}) que também seja uma constante independente da coordenada espacial $x$. Definimos então a função $\Gamma$, tal que
\begin{equation}
	\frac{d\Gamma}{dr}\equiv a\left[2\frac{dY}{dr} - Z\frac{dF}{dr} + 2e^{-F}~\frac{d\psi}{dr} - \frac{dZ}{dr}\right].
\end{equation}
Portanto, escrevemos que
\begin{equation}
	\Gamma= a\left[\left(\int_{r_0}^{r}-ZdF + 2e^{-F}d\psi\right) + 2Y - Z\right], \label{419}
\end{equation}
onde $a$ e $r_0$ são constantes arbitrárias. Porém, podemos determinar um valor para a constante $a$ tal que $\Gamma=M$ e usando $Z=J-2M/\lambda$, assim $a=\lambda/2$.

Com a função $\Gamma$ fazendo o papel da massa do buraco negro acústico, se obtivermos uma expressão para a temperatura Hawking equivalente, podemos formular as leis da termodinâmica para o modelo análogo. Assim, usando a equação (\ref{f38}) para a métrica acústica, obtemos a gravidade superficial acústica
\begin{equation}
	\kappa_a=\left.\frac{1}{2}\frac{dZ}{dr}\right|_{r=r_h}, \label{420}
\end{equation}
e da definição da temperatura Hawking, $T=\kappa/2\pi$, temos que
\begin{equation}
	T_a=\frac{1}{4\pi}\frac{dZ}{dr}=\frac{1}{4\pi}\left[\frac{1}{c_s}\frac{d}{dx}(c_s^2-v_0^2)\right]_{c_s=v_0}, \label{421}
\end{equation}
que é exatamente o que obtemos no capítulo anterior. Podemos escrevê-la também em função das variáveis novas,
\begin{equation}
	T_a=\left[\frac{1}{4\pi\rho_0}\left(2\frac{dY}{dx} - Z\frac{dF}{dx} + 2e^{-F}~\frac{d\psi}{dx}\right)\right]. \label{422}
\end{equation}

Para a massa $M_a$ do buraco negro acústico, tomamos a equação (\ref{419}) calculada no horizonte de eventos com $a=\lambda/2$. Então,
\begin{equation}
	M_a=\frac{\lambda}{2}\left[\int_{x_0}^{x_h}dx\left(-Z\frac{dF}{dx}+2e^{-F}~\frac{d\psi}{dx}\right)+2Y(x_h)\right],\label{423}
\end{equation}
onde usamos que $Z(x_h)=0$. Com isso podemos escrever que
\begin{equation}
	dM_a=T_a(2\pi\lambda\rho_0dx).
\end{equation}
Definindo a função
\begin{equation}
	S_a\equiv 2\pi\lambda\int_{\infty}^{x_h}\rho_0dx, \label{424}
\end{equation}
obtemos finalmente o equivalente a primeira lei da termodinâmica,
\begin{equation}
	dM_a=T_adS_a, \label{425}
\end{equation}
onde $S_a$ é a entropia do buraco negro acústico. É importante notar que, se tivéssemos utilizado diretamente a definição de entropia do sistema gravitacional contido nas equações (\ref{f49}) e (\ref{f50}), teríamos obtido o mesmo resultado para a entropia do sistema acústico. Isso equivale dizer que $S_a$ é equivalente a $S$ até mesmo em sua definição.

\subsection*{Caso - b}
A seguir mostramos que os resultados são exatamente os mesmos quando consideramos o potencial externo como sendo nulo e adotamos a área da seção transversal por onde o fluido escorre como uma função da posição. Neste caso, as equações da dinâmica do fluido escritas em função das variáveis (\ref{414}), são
\begin{align}
	&\frac{dZ}{dr}=2\frac{dY}{dr}-Z\frac{dF}{dr},\nonumber \\
	\nonumber \\
	&\frac{d}{dr}[Y(Y-X)A^2]=0. \label{426}
\end{align}
Da mesma forma como fizemos no caso anterior, introduzimos o vínculo
\begin{equation}
	2\frac{dY}{dr}-Z\frac{dF}{dr}=\lambda V, \label{427}
\end{equation}
e assim as equações (\ref{426}) podem ser feitas equivalentes com a dinâmica gravitacional da equação (\ref{411}). Nesse caso, o vínculo acima também representa uma equação de estado para o fluido se $A(x)$ for dado. Por outro lado, se a equação de estado for conhecida, o vínculo representará uma expressão para $A(x)$.

A função constante que pode ser interpretada como a massa do buraco negro acústico é
\begin{equation}
	\Gamma=a\left[\int_{r_0}^{r}(-ZdF) +2Y - Z\right]. \label{428}
\end{equation}
E desta equação obtemos a massa $M_a$,
\begin{equation}
	M_a=\frac{\lambda}{2}\left[\int_{x_0}^{x_h}dx\left(-Z\frac{dF}{dx}\right) +2Y(x_h)\right], \label{429}
\end{equation}
e a temperatura Hawking acústica
\begin{equation}
	T_a=\frac{1}{4\pi\rho_0}\left[2\frac{dY}{dx}-Z\frac{dF}{dx}\right]_{c_s=v_0}. \label{430}
\end{equation}
Assim a primeira lei da termodinâmica é respeitada com a entropia do sistema acústico dada novamente pela equação (\ref{424}).

%% file: Conclusoes.tex
\setcounter{chapter}{5}
\chapter{Considerações Finais}
\hspace*{\parindent}

Este trabalho apresentou uma revisão teórica sobre um modelo acústico análogo ao buraco negro de Schwarzschild. Explicitamos a maneira como esta analogia pode ser estabelecida quando perturbamos as equações da dinâmica dos fluidos e com isso conseguimos descrever a propagação das ondas sonoras no ferramental matemático da relatividade geral de Einstein. Essas ondas sonoras, que foram tomadas como sendo pertubações no campo de velocidade de propagação do fluido, descrevem geodésicas nulas em um espaço-tempo efetivo governado por uma métrica dependente da velocidade local do som, da densidade e da velocidade do fluido:
\begin{eqnarray}
ds^2=\frac{\rho_0}{c_s}\left[-c^2_sdt^2+\delta_{ij}(dx^i-v_0^idt)( dx^j-v_0^jdt)\right].
\end{eqnarray}

A métrica acústica obtida tem termos cruzados do tipo tempo-espaço e, para efeito de comparação, devemos escrever a métrica de Schwarzschild na forma de Painlevè-Gullstrand (equação \ref{f317}). Notamos que a equivalência entre essas duas métricas é conforme, o que a princípio não apresenta problemas quanto ao uso dessa analogia para o estudo da radiação Hawking visto que a gravidade superficial e, consequentemente, a temperatura Hawking são grandezas invariantes sob transformações conformes da métrica.

Para obter um equivalente acústico da gravidade superficial de um buraco negro procedemos da mesma maneira que foi apresentado no capítulo 2. Através da definição de horizontes de Killing obtém-se uma relação entre a gravidade superficial e a derivada dos elementos da métrica. O resultado obtido é exatamente o mesmo que aqueles apresentados em \cite{unruh, visser}, mas que foram calculados de uma forma distinta.
\begin{equation}
\kappa=\left.\frac{1}{2\rho_0}\frac{d}{dx}\left(-g_{tt}\right)\right|_{v_o=c_s} = \left.\frac{1}{2c_s}\frac{d}{dx}(c_s^2-v_0^2)\right|_{v_o=c_s}.
\end{equation}

Sobre o fato de obtermos uma temperatura Hawking para o buraco negro acústico, equação (\ref{f322}), conclui-se que a radiação Hawking é um fenômeno puramente cinemático já que não utilizamos as equações de campo de Einstein para descrever o nosso modelo \cite{visser4}. Isto é, para uma geometria Lorentziana, basta que haja um horizonte de eventos que a radiação Hawking surgirá sem que seja necessário previamente impor uma dinâmica para o sistema. No entanto, o fato do modelo acústico ter uma métrica conforme e não exatamente análoga à métrica de Schwarzschild, gera complicações se tentarmos imitar a dinâmica da relatividade geral com este modelo. Mais específicamente, uma analogia à entropia de um buraco negro seria bastante complicado pois teríamos de fazer modificações na ação da mecânica dos fluídos, já que esta entropia está diretamente relacionada com o teorema da área do horizonte de eventos de um buraco negro, explicitado na equação (\ref{f49}).

Mesmo assim, essa dificuldade pode ser vencida se trabalharmos em cima de uma analogia a um buraco negro bidimensional. Este sistema 2D é obtido utilizando a descrição do buraco negro de Schwarzschild através de um campo escalar, o dílaton. Com isso, a equivalencia entre as métricas ocorre de forma exata e o sistema acústico é somente bidimensional também, permitindo estender a analogia até o nível dinâmico.

Neste ponto fica claro uma diferença crucial entre os modelos gravitacionais e acústico. A relatividade geral é uma teoria independente da métrica, as equações de campo são derivadas de uma determinada ação não sendo necessário a definição de uma métrica para isso. Já no modelo acústico, a métrica que governa a propagação das ondas sonoras é obtida na pertubação das equações de Euler e da continuidade, ou seja, uma consequência que surge quando descrevemos a propagação do som usando o ferramental matemático da relatividade geral. Isso implica que devemos previamente impor um vínculo no sistema gravitacional fixando uma métrica para ele, neste caso uma métrica no calibre de Schwarzschild.
\begin{equation}
	ds^2=-X(r)dt_s^2 + X(r)^{-1}dr^2.
\end{equation}

Feito isso podemos definir uma função equivalente à massa do buraco negro gravitacional e, a partir dessa função, obter uma entropia para buraco negro acústico assim como a primeira lei da termodinâmica do buraco negro acústico. No entanto este modelo também apresenta suas limitações devido as aproximações utilizadas sobre ele. Essa analogia à dinâmica de um sistema gravitacional só é válida para soluções estáticas, o que nos impede de utilizá-las em várias situações interessantes onde o caso não estático é essencial, por exemplo, uma investigação da evaporação de um buraco negro através do cálculo das reações de fundo, fenômeno que nem mesmo Hawking levou em consideração quando realizou seus cálculos. Também há o fato de não considerarmos um modelo 2D com massa. A ausência de fontes para o campo gravitacional representa um enorme simplificação no modelo pois as equações dinâmicas ficam muito mais simples. Mas, sabemos que é nescessário a presença de fontes para descrevermos uma situação realística, porém, encontrar uma grandeza no sistema acústico similar ao tensor momento-energia do campo gravitacional pode se tornar algo muito complicado de se fazer.

Como uma sequência desta dissertação, pretende-se aplicar simulações numéricas ao estudo da física dos buracos negros acústicos (\cite{fabri1},\cite{fabri2},\cite{fabri3}) não restringindo a análise a uma região muito próxima do horizonte sônico, isto é, a região onde o movimento do fluido muda de subsônico para supersônico, em conexão com modelos de Física da Matéria Condensada. Neste sentido, tal procedimento aplicado  a correntes de fluido (inclusive correntes de fluido em bocais de Laval) pode averiguar a formação de buracos negros acústicos e estimar a contrapartida clássica da radiação Hawking, ou seja, o espectro de potência da perturbação do potencial velocidade do fluido.

%% file: bibliografia.tex
\pagestyle{fancy}                       
\fancyhf{}                              
\renewcommand{\chaptermark}[1]{         
  \markboth{\chaptername\ \thechapter.\ #1}{}} %
\fancyfoot[R]{\footnotesize \thepage}   
\fancyhead[L]{\footnotesize \leftmark}
\renewcommand{\headrulewidth}{0.2pt}    
\addtolength{\headheight}{0.5pt}
\makeatletter 
\def\cleardoublepage{\clearpage\if@twoside \ifodd\c@page\else%
   \hbox{}%
    \thispagestyle{empty}
    \newpage%
    \if@twocolumn\hbox{}\newpage\fi\fi\fi} 
\makeatother